\documentclass[12pt,prd,showpacs,showkeys,superscriptaddress]{revtex4} 
\usepackage{amsfonts}
\usepackage{amssymb}
\usepackage{amsmath}

\usepackage{multirow}
\usepackage{setspace}
\usepackage{tikz}
\usetikzlibrary{shapes,arrows}
\tikzstyle{condition} = [rectangle, draw,text width=6em, minimum height=2em, text badly centered, node distance=4cm, inner sep=0pt]
\tikzstyle{line} = [draw, -latex']
\tikzstyle{type} = [draw, ellipse,node distance=3cm, minimum height=2em]
\usepackage{color}
\definecolor{DarkGreen}{rgb}{0,0.4,0}
\definecolor{DeepBlue}{rgb}{0,0,.4}
\definecolor{deepred}{rgb}{.5,0,0}
\usepackage[naturalnames=true,colorlinks=true,citecolor=DarkGreen,urlcolor=blue,linkcolor=DeepBlue]{hyperref}
\begin{document}

\title{What happens to Petrov classification\\
on horizons of axisymmetric dirty black holes}
\author{I. V. Tanatarov}
\affiliation{Kharkov Institute of Physics and Technology,\\
1 Akademicheskaya, Kharkov 61108, Ukraine}
\affiliation{Department of Physics and Technology, Kharkov V.N. Karazin National
University, 4 Svoboda Square, Kharkov 61077, Ukraine}
\email{igor.tanatarov@gmail.com}
\author{O. B. Zaslavskii}
\email{zaslav@ukr.net}
\affiliation{Department of Physics and Technology, Kharkov V.N. Karazin National
University, 4 Svoboda Square, Kharkov 61077, Ukraine}

\begin{abstract}
We consider axisymmetric stationary dirty black holes with regular non-extremal or extremal horizons, and compute their on-horizon Petrov types. The Petrov type (PT) in the frame of
 the observer crossing the horizon can be different from that formally obtained in the usual (but singular in the horizon limit) frame of an observer on a circular orbit. We call this entity the boosted Petrov type (BPT), as the corresponding frame is obtained by a singular boost from the regular one. The PT off-horizon can be more general than PT on-horizon and that can be more general than the BPT on horizon. This is valid for all regular metrics, irrespective of the extremality of the horizon. We analyze and classify the possible relations between the three characteristics and discuss the nature and features of the underlying singular boost. The three Petrov types can be the same only for space-times of PT D and O off-horizon. The mutual alignment of principal null directions and the generator in the vicinity of the horizon is studied in detail. As an example, we also analyze a special class of metrics with utra-extremal horizons (for which the regularity conditions look different from the general case) and compare their off-horizon and on-horizon algebraic structure in both frames.
%\keywords{Petrov type \and dirty black holes \and horizon}
%\PACS{04.70.Bw \and 97.60.Lf}
\end{abstract}

\maketitle

\newpage\begin{spacing}{1.3}
\tableofcontents
\end{spacing}\newpage

\section{Introduction}

\label{sec:intro} One of the basic results of the theory of gravitation
consists in the possibility to classify different types of gravitational
field in a universal manner. This classical result was obtained by Petrov 
\cite{p} and applies to any metric theory irrespective of its dynamic
contents. It turned out that there are only six algebraic types of the
gravitational field, distinguished by algebraic structure of the Weyl
tensor, namely the multiplicities of its four principal null directions
(pnds).

One of important properties of Petrov classification is that it is
coordinate independent. More specifically, as the classification involves
the use of some null tetrad, the result does not depend on the choice of
this tetrad and the associated observer. Near a black hole horizon, however,
the situation in this regard becomes non-trivial.

Let us consider, for simplicity, a static black hole and a static observer
which resides near its horizon. The static coordinate frame, associated with
such observers, is commonly used in order to study various aspects of
gravity near the horizon \cite{fn}. However, as is well-known, the static
coordinate frame breaks down on the horizon itself. In order to remedy this
situation, one is led to introduce Kruskal-like coordinate frame, that
covers the horizon and its vicinity with some regular coordinate chart. This
frame is attached to an observer who is falling inside the black hole and is
more "physical", as the corresponding metric is regular near the horizon.
This concerns not only black hole space-times themselves but also their
descriptions with the help of time-like surfaces that approach a horizon
only asymptotically (membrane paradigm \cite{m}, quasiblack holes \cite{qbh}).

The reason for the break-down of the static frame is that static observers
do not exist on the horizon: the local Lorentz boost, which relates the
frames of the static observer and the one of arbitrary regular particle
crossing the horizon (we do not consider here the so-called critical
particles, studied in the context of ``black holes as particle
accelerators'' \cite{ban,k,bifurk}), becomes singular in the horizon limit,
as the Lorentz factor diverges. Due to this singular boost the algebraic
structure of various geometric quantities in the singular static frame
becomes highly constrained and they acquire quite specific forms (see \cite{v2,v3,tan}). Indeed, this seems to be the geometric origin of the ``hidden
symmetry'' of black hole horizons, which is related in turn to the universality of Bekenstein-Hawking entropy and to the many miraculous
properties of black hole horizons as seen by a remote or static observer (see e.g. \cite{conf,carlip}).

One of the consequences of the singular Lorentz boost is that in the horizon
limit two vectors, which are not parallel in the regular frame, can seem 
parallel in the static frame. Thus if one attempts to calculate the Petrov
type of a black hole spacetime at a point on horizon using the static frame 
\cite{d}, he obtains incorrect results. This was pointed out in \cite{vo},
where Petrov classification was studied near horizons of generic dirty
static black holes. Let us call the result of the corresponding procedure
(given explicitly in the text below) the ``boosted Petrov type'' (BPT) at
the horizon, in order to differentiate it from the true unique Petrov type
(``regular'' Petrov type, RPT) as calculated in a regular frame at the given
point. Although the two are generally different, this fact in no way implies
that the boosted Petrov type is a characteristic devoid of physical meaning
and should be ignored. On the contrary, the static frame is the one
associated with the Killing vector fields, so that the underlying symmetries
manifest themselves most clearly, and the on-horizon limit is well-defined.
Moreover, as mentioned above, the singular behaviour of the static frame is
directly related to the thermodynamical properties of the horizon and its
features should be carefully studied.

In the present paper we extend the analysis of Ref. \cite{vo} to the case of
generic dirty (surrounded by matter) rotating axially symmetric black holes
and proceed to study the asymptotic structure of pnds in both the regular
and singular frames. Static observers are replaced in stationary spacetimes
by observers on circular orbits. The present symmetries allow one to
simplify significantly the general classification scheme; we also elaborate
on the global structure of double pnds in spacetimes of types N and II. For
the analysis of the near-horizon structure we take advantage of our recent
results \cite{tan} where the near-horizon geometry was extensively
investigated for the two frames considered.

As the horizon is a light-like surface, the effect under discussion resembles to some extent the known fact that some space-times obtained one from another by boosting a black hole metric \cite{boost} can have different Petrov types. However, the feature we discuss concerns not the relation between two different space-times but is inherent to the
submanifold (i.e. horizon) of the same space-time, for different groups of observers (frames). The corresponding limiting procedure therefore turns out to be essentially different from the one implied in \cite{boost}: in particular, it relies on the properties of spacetime not only at the considered point but also in its vicinity (thus is ``less local''), and essentially uses symmetry properties. Therefore the present results are not trivial extension of those for static spacetimes \cite{vo}, since an additional Killing vector changes now the whole picture.  When symmetry is absent, the list of possible relationships between PT and BTP on the horizon is also given.

There is also interest in the relation between the on-horizon limit and the
off-horizon behavior in its vicinity. The Petrov type in a given point
depends on which of some set of algebraic entities (the so-called Weyl
scalars and their combinations) vanish in this point. It does not depend on
the rate with which some of them may approach zero. Therefore, it may happen
that in some cases there is no continuity here, so the Petrov type can
change: on a submanifold of lesser dimensionality (e.g. the horizon) its
Petrov type can in principle be different from that in the bulk space-time
around it \cite{exact}. Thus we discuss here two types of relations: between
the regular and boosted Petrov types on the horizon and between the types
off-horizon and on-horizon. In most textbooks on gravitation it is stressed
that the geometry near a horizon has nothing special, so that an observer
that crosses the horizon cannot detect anything peculiar. Nevertheless, we
show that in some cases measurement of the Petrov type can actually
distinguish the horizon locally.

The paper is organized as follows. In Sec.~\ref{sec:regular}, we briefly
review the regular axisymmetric stationary metrics to be considered, and
show how the frame of the observer crossing the horizon with finite proper
acceleration is constructed. In Sec.~\ref{sec:petrov}, we briefly review the
Petrov classification scheme, adjust it for axisymmetric space-times, and
discuss in some detail the peculiar global properties for Petrov types II
and N. In section~\ref{sec:regularity}, we cover the transformation of key
quantities (Weyl scalars) between the two frames and show how regularity
conditions restrict the regular and boosted algebraic types. We compare and 
distinguish different mathematical procedures describing singular boosts in different situations. 
The restrictions by the off-horizon Petrov types are analyzed in Sec. \ref{sec:Off}. The off-horizon
and on-horizon asymptotic structure of the principal null directions is
discussed in Section~\ref{sec:principal}. As examples, two unusual regular
metrics, explicit expressions for which were obtained in \cite{tan}, are
analyzed in Section~\ref{sec:example}. They are shown to be algebraically
special, asymptotically Ricci flat and flat at the horizon respectively. The
geometry of horizon itself is flat in both cases. The regular and boosted
Petrov types for them are also different, in conformity with the general
scheme, the regular one being more general. We conclude with discussion in
Sec.~\ref{sec:discussion}. 

It is worth stressing, that all calculations are done in the horizon limit,
when the lapse function $N^{2}$ tends to zero and is a small parameter. Thus
we do not use any information about topology and global properties of the
metrics considered (except for the examples), and only investigate their
local properties in the horizon limit.

\section{Regular axisymmetric stationary metrics}

\label{sec:regular} We consider a stationary axially symmetric space-time.
In the vicinity of the horizon its metric can always be written in Gaussian
normal coordinates (see \cite{v2}): 
\begin{equation}  \label{metric-n}
ds^{2}=-N^{2}dt^{2}+g_{\phi\phi}(d\phi-\omega dt)^{2} +dn^{2}+g_{zz}dz^{2},
\end{equation}
where $n$ is the proper distance to the horizon, and all metric functions
depend on $n$ and $z$ only; on the horizon $N^{2}\to 0$. The coordinate $n$,
however, is badly-behaved in the vicinity of the horizon, since it is
applicable in the outer region only and cannot be used inside the horizon,
and it is often more convenient to use the "quasiglobal" radial coordinate $%
r(n)$, defined so that 
\begin{equation}  \label{metric-g}
ds^{2}=-N^{2}dt^{2}+g_{\phi\phi}(d\phi-\omega dt)^{2} +\frac{dr^2}{A(r)}%
+g_{zz}dz^{2},
\end{equation}
with $A(r)\sim N^{2}$ as $r\to 0$, where we set $r=0$ at the horizon. Note
that it is different from the Boyer-Lindquist coordinate $\tilde{r}$ for the
Kerr-Newman metric, as $g_{\tilde{r}\tilde{r}}$ depends also on $z$.

If $N^{2}\sim r^{p}$, with $p\in\mathbb{N}$, the horizon is said to be of
extremality $p$: $p=1$ corresponds to non-extremal horizons, $p=2$ to
extremal ones and $p\geq 3$ to ultra-extremal ones.

\subsection{Orbital (OO) and falling (FO) observers}

In a stationary space-time the most simple class of observers, the analogue
of static observers in a static space-time, are OZAMOs: orbital zero angular
momentum observers. They orbit the black hole on a circular trajectory $%
r=const$, $z=const$, have angular momentum $L$ equal to zero, and are
usually referred to as ZAMOs \cite{72}. For brevity, we will also call them "orbital
observers", OOs. The OO tetrad $\{e_{(i)}\}$, with $i=0,1,2,3\equiv
t,\phi,r,z$, is 
\begin{align}
e_{(0)}&=-Ndt,  \label{OZAMO-0} \\
e_{(1)}&=\sqrt{g_{\phi\phi}}\;(d\phi -\omega dt); \\
e_{(2)}&=dn=A^{-1/2}dr; \\
e_{(3)}&=\sqrt{g_{zz}}\;dz.  \label{OZAMO-3}
\end{align}

The OZAMOs, by definition, do not cross the horizon and thus are inefficient
in probing the horizon limit. So let us construct the FZAMOs, falling zero
angular momentum observers, by constructing the corresponding tetrad $%
\{f_{(i)}\}$:

\begin{enumerate}
\item first, we take the OZAMO's tetrad $\{e_{(i)}\}$;

\item rotate it in the $(r-z)$ plane by angle $\theta$ thus obtaining new
tetrad $\{h_{i}\}$: $h_{(0,1)}=e_{(0,1)}$ and 
\begin{equation}
\Bigg(\!\!%
\begin{array}{c}
h_{(2)} \\ 
h_{(3)}%
\end{array}%
\!\!\Bigg)= \Bigg(%
\begin{array}{cc}
\cos\theta & \sin\theta \\ 
-\sin\theta & \cos\theta%
\end{array}%
\Bigg) \cdot \Bigg(\!\!%
\begin{array}{c}
e_{(2)} \\ 
e_{(3)}%
\end{array}%
\!\!\Bigg);
\end{equation}

\item add Lorentz boost along the new $(-\hat{r})$ direction (towards the
horizon) with Lorentz factor $\gamma=(1-v^2)^{-1/2}$, thus obtaining $%
\{f_{(i)}\}$: for the basis 1-forms $f_{(1,3)}=h_{(1,3)}$ and 
\begin{equation}
\Bigg(\!\!%
\begin{array}{c}
f_{(0)} \\ 
f_{(2)}%
\end{array}%
\!\!\Bigg) = \gamma \Bigg(%
\begin{array}{cc}
1 & v \\ 
v & 1%
\end{array}%
\Bigg) \cdot \Bigg(\!\!%
\begin{array}{c}
h_{(0)} \\ 
h_{(2)}%
\end{array}%
\!\!\Bigg).
\end{equation}
\end{enumerate}

The relative Lorentz factor of OZAMO and FZAMO is equal to 
\begin{equation}
\gamma=-f_{(0)}\cdot e_{(0)}=\frac{E}{N},
\end{equation}
where $E$ is the FZAMO's energy per unit mass (integral of motion), and in
the horizon limit $\gamma$ always diverges, unless the falling particle is
fine-tuned, with $E\to 0$. This is just the indication that the OZAMO's
frame breaks down at the horizon, not that something is wrong with FZAMO.
The mentioned fine-tuned particles are a special case ($L=0$) of the
so-called critical particles, which in general obey $E-\omega L \sim N$ in
the horizon limit. For them the proper time of reaching the horizon is
infinite, and their existence leads to such phenomena as the BSW effect \cite%
{ban}. The consideration below works only for non-critical observers.

A valid test particle with 4-velocity $u^\mu=f_{(0)}^{\mu}$ must have smooth
worldline, or equivalently, its acceleration scalar 
\begin{equation}
a=\sqrt{a^\mu a_\mu},\quad a^{\mu} =u^{\nu}\nabla_{\nu}u^{\mu}
\end{equation}
should be finite. Moreover, all four tetrad components of acceleration in the proper frame of the particle should be finite too \cite{bsw-force}. 

Note now, that in the presence of a Killing vector $\xi^\mu$ (there are two
in our case, $\partial_t$ and $\partial_\phi$), the conservation of the
corresponsing integral $u^\mu \xi_\mu$ along the worldline of the particle
is equivalent to the projection of its acceleration on this Killing vector
being zero, as simple manipulations show that 
\begin{equation*}
a^\nu \xi_\nu =u^\mu \nabla_\mu (u^\nu \xi_\nu).
\end{equation*}
The angular momentum of an FZAMO by construction is zero and conserved, so $%
a_\phi =0$. By taking $E=const$, we also get $a_t =0$. Then both tetrad
components are also equal to zero. As
this is true along the worldline, not only in the horizon limit, it is not
affected by the breakdown of the coordinate frame in which the initial
projection was calculated: when multiplying identical zero by a divergent
quantity one still obtains zero.

Given these restrictions, finiteness of $a$ is equivalent to boundedness of $r$ and $z$ 
tetrad components of acceleration in the OZAMO frame, and these two conditions determine valid observers. It can be shown that those are satisfied if $\theta=O(N)$: the
particle is crossing the horizon "vertically" relative to OO. In terms of
tetrad components of physical velocity relative to OZAMO $\{\hat{v}_{(i)}\}$
this implies $\hat{v}_{z}=O(N) $ and $\hat{v}_{r}=1-o(1)$, while zero
angular momentum means $\hat{v}_{\phi}=0$. Moreover, if the horizon is
non-extremal or the geometry is spherically symmetric, this condition also
turns out to be necessary (it is well-known that "the rain falls down
vertically for ZAMOs" (see \cite{TaylorWheeler} for example); for
non-extremal horizons this, however, turns out to be true for any real
particle, not necessarily in free fall). We will not go into more details
here, as this should be reported elsewhere. In the more general case of
extremal axisymmetric horizons this is not so, but the particles with
non-zero on-horizon value of $\theta$ can only possibly exist in a special
class of metrics, which, as opposed to the exotic metrics discussed below,
are not even regular in the considered sense, and so will also not be
addressed here.

Thus, by taking $E=const$ and $\theta=O(N)$, we single out FZAMOs with
finite proper acceleration, which can be verified easily by hand. Those are
valid observers crossing the horizon, which we will be using for probing the
metric in this limit. This class of observers contains, by construction,
both true geodesic observers and the ones with $z=const$ (which are
different), considered in \cite{vo}. The later represent the most simple
subclass of FZAMOs, which we will denote FOs, for "falling observers".

\subsection{Regularity conditions}

We consider here only metrics with regular horizons, excluding the so-called
"truly naked black holes" \cite{vo} with non-scalar curvature singularities
at the horizon. A non-scalar singularity is said to be present when scalar
curvature invariants are finite but some tetrad components of the curvature
tensor in a tetrad attached to an observer crossing the horizon~-- an
FZAMO~-- diverge \cite{el}. Physically, this would mean divergent tidal
forces that this observer experiences.

The corresponding regularity conditions in terms of expansions of the metric
functions near the horizon were obtained explicitly in \cite{tan}. They
imply that all regular metrics of the considered class with horizons of any
extremality are divided into two classes. The metrics of the first one,
which we will call generic regular metrics hereafter, obey two simple
conditions of "rigidity" (the second condition can also be called this way,
as it implies the "rigidity", i.e. slow enough variation, of the lapse
function) 
\begin{equation}  \label{GenericRegCond}
\partial_{z}\omega=O(N^2),\quad \partial_{z}\ln N^2 =O(N^2).
\end{equation}
Regular metrics with non-extremal (single) or extremal (double) horizons can
only be generic. There are, however, regular metrics with ultra-extremal
horizons that are not generic, and are therefore, by definition, exotic.
Those obey a weaker set of six conditions (Eqs. (76-81) of \cite{tan}),
while the two conditions (\ref{GenericRegCond}) do not hold. In particular,
for triple horizons $p=3$ there are only two exotic metrics, explicit
expansions for which were obtained in \cite{tan} and will be given below.

Below we will not need regularity conditions in any explicit form. We will
only use the condition that Weyl tensor tetrad components in the FO frame
are finite; also we will use the result of \cite{tan}, that tetrad
components of curvature (and thus Weyl) tensor in the OO frame are bounded
as long as curvature invariants are finite. Thus if the the tetrad
components of Weyl tensor are finite in the FO frame, they are finite in
both.

\section{Petrov classification: axisymmetric spacetimes}

\label{sec:petrov}

\subsection{Petrov classification}

\label{subsec:petrov}

In order to determine the metric's Petrov type we use the classic scheme,
detailed in e.g. \cite{exact} or \cite{Gr}, which we repeat here briefly for
consistency. Given an orthonormal frame attached to the observer in question 
$\{e_{(i)}\}$, we construct the null complex tetrad $\{l_{(i)}\}=%
\{l_+,l_-,m_+,m_-\}$: 
\begin{align}
l_{\pm}&=\frac{e_{(0)}\pm e_{(2)}}{\sqrt{2}};  \label{null-02} \\
m_{\pm}&=\frac{e_{(1)}\pm i e_{(3)}}{\sqrt{2}},  \label{null-13}
\end{align}
such that metric tensor in this tetrad has the form 
\begin{equation}
g_{(i)(j)}^{(null)}=g_{\mu\nu}l_{(i)}^{\mu}l_{(i)}^{\nu} =\bordermatrix{
&l_{+}&l_{-}&m_{+}&m_{-}\cr l_{+}&0 & -1 & 0 & 0 \cr l_{-}&-1 & 0 & 0 & 0
\cr m_{+}&0 & 0 & 0 & +1 \cr m_{-}&0 & 0 & +1 & 0}.
\end{equation}
The five Weyl scalars (or Cartan scalars) then are the independent
components of the Weyl tensor $C_{\mu\nu\rho\sigma}$ in this tetrad: 
\begin{align}
\Psi_{0}&\equiv C_{(1)(3)(1)(3)} =C_{\alpha\beta\gamma\delta}\;
l_{+}^{\alpha}m_{+}^{\beta} l_{+}^{\gamma}m_{+}^{\delta}; \\
\Psi_{1}&\equiv C_{(1)(3)(1)(2)} =C_{\alpha\beta\gamma\delta}\;
l_{+}^{\alpha}m_{+}^{\beta} l_{+}^{\gamma}l_{-}^{\delta}; \\
\Psi_{2}&\equiv -C_{(1)(3)(2)(4)} =-C_{\alpha\beta\gamma\delta}\;
l_{+}^{\alpha}m_{+}^{\beta} l_{-}^{\gamma}m_{-}^{\delta}; \\
\Psi_{3}&\equiv C_{(1)(2)(4)(2)} =C_{\alpha\beta\gamma\delta}\;
l_{+}^{\alpha}l_{-}^{\beta} m_{-}^{\gamma}l_{-}^{\delta}; \\
\Psi_{4}&\equiv C_{(2)(4)(2)(4)} =C_{\alpha\beta\gamma\delta}\;
l_{-}^{\alpha}m_{-}^{\beta} l_{-}^{\gamma}m_{-}^{\delta}.
\end{align}

A vector $k$ is said to be a principal null direction (pnd), if 
\begin{equation}  \label{pndK}
k_{[\mu}C_{\alpha]\beta\gamma[\delta} k_{\nu]}k^{\beta}k^{\gamma}=0.
\end{equation}
There are four such vectors. The first vector of the tetrad, $l_{+}$, is a
pnd if and only if $\Psi_{0}=0$.

When applying a "null rotation" of the tetrad, with fixed $l_{-}$ and
complex parameter $\lambda$ 
\begin{equation}  \label{nullrot-vec}
l_{+}^{\prime}=l_{+}+\lambda m_{-}+\bar{\lambda} m_{+}
+|\lambda|^{2}l_{-},\quad l_{-}^{\prime}=l_{-}, \quad
m_{+}^{\prime}=m_{+}+\lambda l_{-}
\end{equation}
(overbar denotes complex conjugation), the scalar $\Psi_{0}$ transforms as 
\begin{equation}  \label{nullrot}
\Psi_{0}^{\prime}=\Psi_{0}-4\lambda\Psi_{1}+6\lambda^{2}\Psi_{2}
-4\lambda^{3}\Psi_{3}+\lambda^{4}\Psi_{4}.
\end{equation}
The four algebraic roots of equation $\Psi^{\prime}=0$ with regard to $%
\lambda$ thus determine the four pnds. If the degree of Eq. (\ref{nullrot})
is $m<4$, then $l_{-}$ is a pnd of multiplicity $4-m$. The root $%
\lambda_{i}=0$ corresponds to $l_{+}$ and formally one can write that $%
\lambda_{i}=\infty$ corresponds to $l_{-}$.

\begin{figure}[ht]
\centering
\begin{tikzpicture}[node distance = 6em, auto]
\pgfsetlinewidth{0.2ex}
    % Place nodes
	\node [condition] (Delta) {$\Delta=0$};
	\node at (4,-0.7) [condition] (IJ) {$I=J=0$};
	\node [condition, below of=IJ, node distance=1.7cm] (KN) {$K=M=0$};
	\node at (8,-2) [condition] (KL) {$K=L=0$};
	\node at (0,-4.3) [type] (I) {$I$};
	\node at (3,-4.3) [type] (II) {$II$};
	\node at (5,-4.3) [type] (D) {$D$};
	\node at (7,-4.3) [type] (III) {$III$};
	\node at (9,-4.3) [type] (N) {$N$};
    % Draw edges
	\path [line] (Delta) -- node {no} (I);
%	\path [line] (Delta) -- (I);
	\path [line] (Delta) -- node {yes} (IJ);
	\path [line] (IJ) -- node {no} (KN);
	\path [line] (IJ) -- node {yes} (KL);
	\path [line] (KN) -- node[swap] {no} (II);
	\path [line] (KN) -- node {yes} (D);
	\path [line] (KL) -- node[swap] {no} (III);
	\path [line] (KL) -- node {yes} (N);
%    \path [line] (decide) -| node [near start] {yes} (update);
%    \path [line,dashed] (system) |- (evaluate);
\end{tikzpicture}

\caption{The flow diagram for determining the Petrov type by the classical
method \protect\cite{exact}.}
\label{TIKZ-Petrov}
\end{figure}
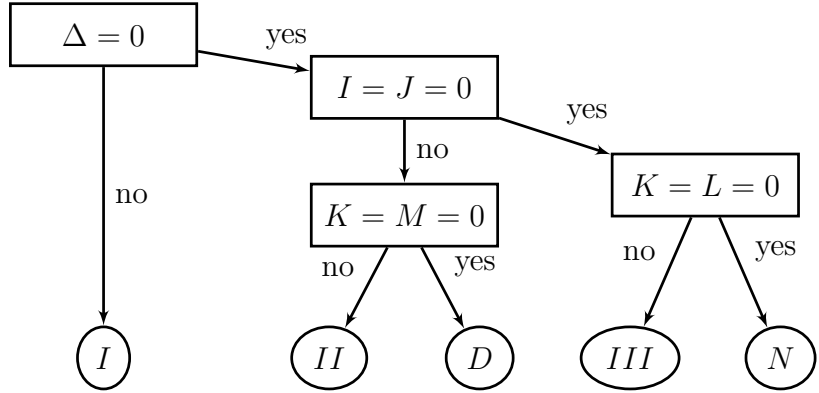

The multiplicities of the pnds determine the Petrov type: type I is
algebraically general, with four distinct pnds. If one of the pnds is double
then the Petrov type is II, two double pnds correspond to type D, a triple
pnd gives type III and a quadruple one type N. If all $\Psi_{i}$ are zero,
the Petrov type is O, the Weyl tensor vanishes and the metric is conformally
flat. Thus one has to analyze the structure of the quartic polynomial $%
\Psi_{0}^{\prime}=0$ (\ref{nullrot}). This is in general done with the help
of the following invariant combinations, called Weyl invariants hereafter: 
\begin{align}
I&=\Psi_{0}\Psi_{4}-4\Psi_{1}\Psi_{3}+3\Psi_{2}^{2};  \label{Inv-I} \\
J&=%
\begin{vmatrix}
\Psi_{0} & \Psi_{1} & \Psi_{2} \\ 
\Psi_{1} & \Psi_{2} & \Psi_{3} \\ 
\Psi_{2} & \Psi_{3} & \Psi_{4}%
\end{vmatrix}%
; \\
K&=\Psi_{1}\Psi_{4}^{2}-3\Psi_{2}\Psi_{3}\Psi_{4} +2\Psi_{3}^{3};
\label{Inv-K} \\
L&=\Psi_{2}\Psi_{4}-\Psi_{3}^{2}; \\
M&=12L^{2}-\Psi_{4}^{2}I;  \label{Inv-M} \\
\Delta&=I^{3}-27J^{2}.  \label{Inv-D}
\end{align}
The flow diagram is shown on Fig. \ref{TIKZ-Petrov}.

\subsection{Axisymmetric stationary space-times}

\label{Petrov-axi-section} Due to axial symmetry, the metric (\ref{metric-g}%
) and all the functions derived from it, including the Weyl scalars, are
invariant under substitution 
\begin{equation}
t\to (-t),\qquad \phi\to (-\phi),  \label{flip}
\end{equation}
which is equivalent to 
\begin{equation}
l_{\pm}\to (-l_{\mp}),\qquad m_{\pm}\to (-m_{\mp}).  \label{flip-null}
\end{equation}
For the chosen tetrad then 
\begin{equation}  \label{OZAMO-symmetry}
\Psi_{0}=\Psi_{4},\qquad\Psi_{1}=\Psi_{3},
\end{equation}
and the general expressions for the invariants (\ref{Inv-I})-(\ref{Inv-D})
are reduced to: 
\begin{align}  
I&=\Psi_{0}^2-4\Psi_{1}^2+3\Psi_{2}^{2}; \label{Iaxi}\\
J&=(\Psi_{0}-\Psi_{2})\big[ \Psi_{2}^2 +\Psi_{0}\Psi_{2}-2\Psi_{1}^2 \big]; \label{Jaxi}
\\
K&=\Psi_{1}\big[2\Psi_{1}^{2}+\Psi_{0}^{2} -3\Psi_{0}\Psi_{2}\big]; \label{Kaxi}\\
L&=\Psi_{0}\Psi_{2}-\Psi_{1}^{2}; \label{Laxi}\\
M&=12L^{2}-\Psi_{0}^{2}I; \label{Maxi}\\
\Delta&= \big(\Psi_{0}^2 +2 \Psi_{1}^2 -3\Psi_{0}\Psi_{2}\big)^{2} \big\{%
(\Psi_{0}+3\Psi_{2})^{2}-16\Psi_{1}^{2}\big\}. \label{Daxi}
\end{align}

Thus the general flow diagram, shown on Fig. \ref{TIKZ-Petrov}, can be
significantly simplified. For the metric to be algebraically special, of at
least Petrov type II, one of the two conditions must be met (type O is
excluded here)

\begin{enumerate}
\item $\Psi_{0}^{2}+2\Psi_{1}^{2}-3\Psi_{0}\Psi_{2}=0$. Eliminating $%
\Psi_{1}^{2}$, we see that 
\begin{equation}
I=3(\Psi_{0}-\Psi_{2})^{2},\quad K=M=0.
\end{equation}
So $I=J=0$ is equivalent to $\Psi_{0}=\Psi_{2}$. Using the general scheme,
we conclude that

\begin{enumerate}
\item if $\Psi_{0}=\Psi_{2}$, then $L=0$ and the Petrov type is N;

\item otherwise the Petrov type is D.
\end{enumerate}

\item $\pm 4\Psi_{1}=\Psi_{0}+3\Psi_{2}$. Then 
\begin{align}  \label{IK-simple}
&I=\tfrac{3}{4}(\Psi_{0}-\Psi_{2})^{2};\quad K=\tfrac{9}{8}%
\Psi_{1}(\Psi_{0}-\Psi_{2})^{2}.
\end{align}
Again $I=J=0$ is equivalent to $\Psi_{0}=\Psi_{2}$, so there are two
variants:

\begin{enumerate}
\item if $\Psi_{0}=\Psi_{2}$, then $\Psi_{0}=\Psi_{2}=\pm\Psi_{1}$, so $%
I,J,K,L,M=0$ and the Petrov type is N;

\item otherwise $K=0$ is equivalent to $\Psi_{1}=0$, which also implies $M=0$%
, and thus

\begin{enumerate}
\item if $\Psi_{1}=0$ then Petrov type is $D$;

\item otherwise the Petrov type is II.
\end{enumerate}
\end{enumerate}
\end{enumerate}

\begin{figure}[ht]
\centering
\begin{tikzpicture} [auto]
\pgfsetlinewidth{0.2ex}
	\node [condition, text width=10em] (lin)
		{$\pm 4\Psi_{1}=\Psi_{0}+3\Psi_{2}$};
	\node [draw, diamond, minimum height=2em, right of=lin, node distance=9em] (or) {OR};
	\node [condition, text width=10em, right of=or, node distance=9em] (quad) {$2\Psi_{1}^{2}=3\Psi_{0}\Psi_{2}-\Psi_{0}^{2}$};
	\draw [-] (lin) -- (or); 	\draw [-] (quad) -- (or);
	\node [type, below of=or, node distance=4em] (I) {I};
	\path [line] (or) -- node{no} (I);
	\node [condition, text width=5em, below of=lin, node distance=5em] (02l) {$\Psi_{0}=\Psi_{2}$};
	\node [condition, text width=5em, below of=quad, node distance=5em] (02q) {$\Psi_{0}=\Psi_{2}$};
	\path [line] (lin) -- node[swap]{yes} (02l);
	\path [line] (quad) -- node{yes} (02q);
	\node [condition, text width=4em, below of=I, node distance=3em] (1a) {$\Psi_{1}=0$};
	\path [line] (02l) -- node {yes} (1a);
	\path [line] (02q) -- node [swap] {yes} (1a);
	\node [type, below of=1a, xshift=-1cm, node distance=4em] (N) {N$^*$};
	\node [type, below of=1a, xshift=1cm, node distance=4em] (O) {O};
	\path [line] (1a) -- node[swap, near end] {no} (N);
	\path [line] (1a) -- node [near end] {yes} (O);
	\node [condition, text width=4em, below of=02l, node distance=4em] (1b) {$\Psi_{1}=0$};
	\node [type, left of=1b, node distance=6em] (II) {II$^*$};
	\node [type, below of=02q, node distance=7em] (D) {D};
	\node [type, below of=1b, node distance=4em] (Dl) {D$^*$};
	\path [line] (02l) -- node [swap] {no} (1b);
	\path [line] (02q) -- node {no} (D);
	\path [line] (1b) -- node {no} (II);
	\path [line] (1b) -- node [swap]{yes} (Dl);
%.. controls +(right:10em) and +(down:4em)..  
\end{tikzpicture}

\caption{The variant of flow diagram for determining Petrov type (Fig.~%
\protect\ref{TIKZ-Petrov}) for axisymmetric stationary space-time in the OO
frame; $^*$asterisks mark the types ruled out in the "generic" case (see
text).}
\label{Petrov-Axi}
\end{figure}
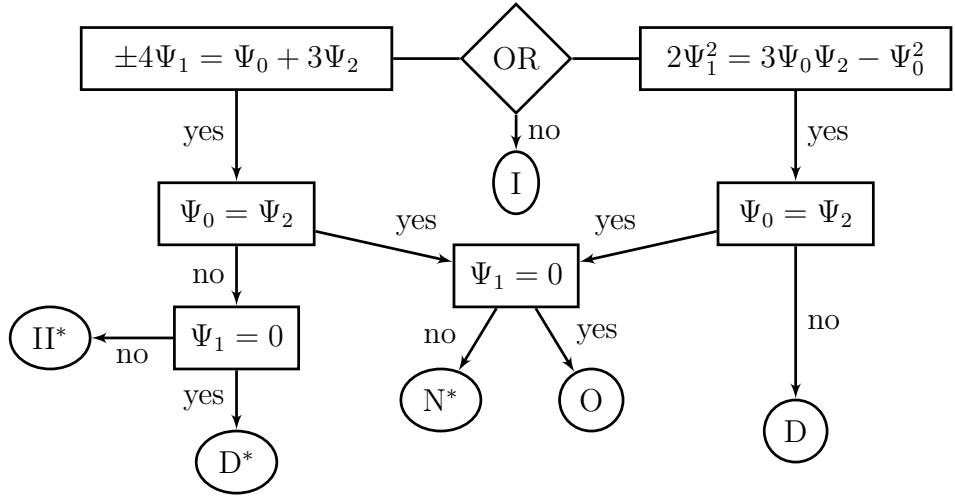

Thus there are no axisymmetric stationary metrics of type III, in agreement
with the known result (\cite{exact}, p.606), and the simplified flow diagram
is shown on Fig.~\ref{Petrov-Axi}.

\subsection{Double circular pnds}

\label{Sec:IIN}

It was shown in \cite{PPO2007} that axial symmetry provides further
restrictions on the algebraic type: namely, type N and II spacetimes have
non-expanding null congruences associated with double null pnds. This is a
specific structure, which cannot be expected to hold true generally, and so
in the ``generic'' case of ``expanding'' spacetimes we are left with Petrov
types I, D and O.

Let us consider the argument in more detail. Axial symmetry implies that
everything is invariant under ``flipping'' (\ref{flip}), which is equivalent
to (\ref{flip-null}). Then the ``flipping'' maps a given pnd also to a pnd
with the same $\lambda$ and multiplicity. Therefore either all pnds are
arranged into pairs of equal mutiplicity (this is possible only for Petrov
type I, D and O metrics) or some of the pnds are mapped onto themselves.

Suppose there is a pnd aligned along $k=(k^{t},k^{\phi},k^{r},k^{z})$ with
multiplicity $m$. Then there is a pnd aligned along $k^{\prime}=(-k^{t},-k^{%
\phi},k^{r},k^{z})$ with the same multiplicity. The two vectors correspond
to different directions unless $k^{r}=k^{z}=0$, which would imply $%
k^{\prime}=-k$. However, if it is the same direction, the same symmetry
implies that the divergence scalars for the associated congruences $%
k^{\mu}_{;\mu}$ and ${k^{\prime}}^{\mu}_{;\mu}=-k^{\mu}_{;\mu}$ are also
equal, and thus $k^{\mu}_{;\mu}=0$. For ``generic'', ``expanding''
spacetimes this would not hold, so the pnds must be broken into pairs of
equal multiplicity. This rules out types III (already seen above from more
general considerations), II and N, which are therefore marked by asterisks
in the flow diagram \ref{Petrov-Axi} and tables below.

In this section we consider the non-expanding case, when a (multiple) pnd is
mapped onto itself, and show explicitly, by algebraic construction, that the
spatial component of such a pnd must be purely azimuthal. This concerns all
spacetimes of type II and N and some of type D.

Suppose there is a given vector $k$, which defines some pnd and thus obeys (%
\ref{nullrot-vec}) with some $\lambda$ 
\begin{align}
k&=l_{+}+\lambda m_{-}+\bar{\lambda}m_{+} +|\lambda|^{2}l_{-},
\end{align}
and ``flipping'' (\ref{flip-null}) maps it to 
\begin{align}
k^{\prime}&=-l_{-}-\lambda m_{+}-\bar{\lambda}m_{-} -|\lambda|^{2}l_{+},
\end{align}
so that 
\begin{align}
-\frac{k^{\prime}}{|\lambda|^{2}}&= l_{+}+\frac{1}{\lambda}m_{-} +\frac{1}{%
\bar{\lambda}}m_{+} +\frac{1}{|\lambda|^2}l_{-}.
\end{align}
The two vectors $k$ and $k^{\prime}$ are parallel, and thus correspond to
the same principal null direction, if and only if $\lambda=1$ or $\lambda=-1$%
. Then using (\ref{nullrot-vec}), we obtain 
\begin{equation}  \label{lambdaPM1}
\lambda=\pm1\quad\Rightarrow\quad k=\sqrt{2}\;\big(e_{(0)}\pm e_{(1)}\big),
\end{equation}
and the integral curves associated with such pnds are null curves $r=const$, 
$z=const$, passing though every point. In the section $t=const$ they are
closed circles.

Note that if $+1$ or $-1$ is a root of 
\begin{equation}  \label{EqAxiSymm}
\Psi_{0}\lambda^{4}-4\Psi_{1}\lambda^{3} +6\Psi_{2}\lambda^{2}
-4\Psi_{1}\lambda +\Psi_{0}=0,
\end{equation}
then it is also a double root (as the equation is invariant under $%
\lambda\to 1/\lambda$), so the pnd that is mapped onto itself is always a
double one. Note, that this explains why there are no axisymmetric
stationary metrics of type III: the triple pnd would have to be mapped onto
itself, implying that $+1$ or $-1$ is a triple root, even though it can only
be of even multiplicity. For $+1$ or $-1$ to be a root (and thus double
root) of Eq. (\ref{EqAxiSymm}), the necessary and sufficient condition is 
\begin{equation}  \label{PsiLinear}
\Psi_{0}\mp 4\Psi_{1}+3\Psi_{2}=0,
\end{equation}
which is exactly the condition obtained above, that must hold for type II
and type N metrics but also allows type D. Thus the considered cases are all
represented in the left half of flow diagram~\ref{Petrov-Axi}). The equation
for the remaining two roots $\lambda_{3,4}$ is then reduced to 
\begin{equation}  \label{roots34}
\lambda^{2}-2\Big(2\frac{\Psi_1}{\Psi_0}\mp1\Big) \lambda+1=0.
\end{equation}
Thus if $\Psi_{1}=0$, then $\lambda_{1,2,3,4}=1,1,-1,-1$ and the type is D;
if $\pm\Psi_{1}=\Psi_{0}=\Psi_{2}$ then $\pm1$ is a quadruple root and the
type is N; otherwise $\lambda_{3,4}$ are different and the type is II.

\section{Regularity and on-horizon algebraic structure}

\label{sec:regularity}

\subsection{Transition between the OO and FO frames}

Suppose the Weyl scalars in the OZAMO frame are $\Psi_{i}$. Let us calculate
the Weyl scalars in the frame of falling observers (FO): the most simple
case of FZAMOs with $\theta=0$. They will be denoted $\Phi_{i}$. Then the
two frames are related by a single Lorentz boost with $\gamma=E/N$, which tends to infinity in the horizon limit.
The 1-forms of the FO null frame $\{\tilde{l}_{i}\}$ 
\begin{align}
\tilde{l}_{\pm}&=\frac{f_{(0)}\pm f_{(2)}}{\sqrt{2}};  \label{Fnull-02} \\
\tilde{m}_{\pm}&=\frac{f_{(1)}\pm i f_{(3)}}{\sqrt{2}},  \label{Fnull-13}
\end{align}
are expressed through the OO null frame $\{l_{i}\}$ (\ref{null-02},\ref%
{null-13}) as 
\begin{equation}
\tilde{l}_{\pm}=\gamma (1\pm v)l_{\pm},\qquad \tilde{m}_{\pm}=m_{\pm},
\end{equation}
or if we introduce $x=\gamma (1+v)\to \infty$, 
%($x\equiv z^{-1}$ where $z$ is from [Pravda Zaslavskii 2005]), 
we get 
\begin{equation}
\tilde{l}_{\pm}=x^{\pm 1} l_{\pm},\qquad \tilde{m}_{\pm}=m_{\pm},
\end{equation}

For the vectors then $\tilde{l}_{\pm }^{\mu }=x^{\mp 1}l_{\pm }^{\mu }$ and
some elementary algebra leads to 
\begin{equation}
\Phi _{0}=x^{-2}\;\Psi _{0};\quad \Phi _{1}=x^{-1}\;\Psi _{1};\quad \Phi
_{2}=1\cdot \Psi _{2};\quad \Phi _{3}=x^{+1}\Psi _{3};\quad \Phi
_{4}=x^{+2}\Psi _{4}.  \label{PhiS}
\end{equation}%
The negative powers of $x$ in these formulas are sometimes called "boost
weights" of the corresponding quantities \cite{vo}.

In the horizon limit the first vectors of both tetrads, $l_{+}$ and $\tilde{l%
}_{+}$, become aligned with the generators of the (future) horizon.
Likewise, near the past horizon the vectors $l_{-},\tilde{l}_{-}$ would be
the counterparts of $l_{+},\tilde{l}_{+}$ in the sense that they are
directed there along the generator of the past horizon.

\subsection{Restrictions imposed by regularity}

The strongest regularity conditions that we use imply that all the tetrad
components of the Riemann tensor, and therefore Weyl tensor as well, remain
finite in the FZAMO frame. 
%\footnote{As this is true for real tetrad components, this is also true for the complex null tetrad, as the components in the two frames are expressed through each other with coefficients $\sim 1/\sqrt{2}^{k}$.}. 
Then the tetrad components in the OZAMO frame are also finite, as the
regularity conditions in the OZAMO frame are weaker \cite{tan}.

As $\Phi_{3,4}$ must be finite, while $x\sim\gamma\sim 1/N\to \infty$, we
have 
\begin{equation}  \label{Psi01-OZAMO}
\Psi_{4}=O(N^{2}),\quad \Psi_{3}=O(N).
\end{equation}
On the other hand, for $\Psi_{0,1}$ to be finite, likewise 
\begin{equation}  \label{Psi34-FZAMO}
\Phi_{0}=O(N^2),\quad \Phi_{1}=O(N).
\end{equation}

Let us define new functions $\psi_{0,1,2}$, such that, taking into account
the symmetry (\ref{OZAMO-symmetry}), 
\begin{align}  \label{psi}
\Psi_{0}=\Psi_{4}=\psi_{0}N^{2},\quad \Psi_{1}=\Psi_{3}=\psi_{1}N,\quad
\Psi_{2}=\psi_{2},
\end{align}
where $\psi_{i}$ are finite due to regularity constraints, but not
necessarily small: 
\begin{equation}  \label{psiO}
\psi_{0},\psi_{1},\psi_{2}=O(1).
\end{equation}

Finally, using 
\begin{equation}  \label{x}
x=\gamma(1+v)\approx \frac{2E}{N},
\end{equation}
for the Weyl scalars in the FO frame we obtain 
\begin{align}
\Phi_{0}&\approx\frac{\psi_{0}}{4E^{2}}N^{4} \sim N^{4}\to 0;  \label{Phi0}
\\
\Phi_{1}&\approx\frac{\psi_{1}}{2E}N^{2} \sim N^{2}\to 0;  \label{Phi1} \\
\Phi_{2}&=\psi_{2}; \\
\Phi_{3}&\approx 2E\psi_{1}; \\
\Phi_{4}&\approx 4E^{2}\psi_{0}.  \label{Phi4}
\end{align}
Their structure in the OO and FO frames is clearly very different. In the OO
frame $\Psi_{0,1,3,4}\to 0$, which means that $l_{\pm}$ are both associated
with double principal null directions; in the FO frame only $\Phi_{0,1}\to0$%
, which means that only $l_{+}$, being aligned with the generator,
corresponds to a double pnd.

It is worth noting, that regularity directly forces $\Psi_{3,4}$ to vanish
on the horizon, while symmetry extends this to $\Psi_{0,1}$, both for
axisymmetric stationary space-times considered here and for static ones \cite%
{vo}. In contrast, in the approach developed for isolated horizons \cite%
{Ashtekar}, zero expansion for any null normal to the horizon leads to
vanishing of $\Psi_{0,1}$.

\subsection{Petrov types at the horizon}

\paragraph{Boosted Petrov type (orbital observer's frame).}

Due to symmetry and regularity we saw that $\Psi_{0,1,3,4}\to 0$ (\ref{psi}%
), so the equation for $\lambda$ in the horizon limit is reduced to $\psi_2
\lambda^2 =0$. Therefore in case $\psi_2$ is separated from zero, we have $%
\lambda=0,0,\infty,\infty$ and two double roots, so the boosted Petrov type
is D. Otherwise all the Weyl tensor components in the OO frame vanish and
the boosted Petrov type is O. The same can be easily shown in terms of the
invariants (\ref{Iaxi}-\ref{Daxi}), and this is the same situation as in the
static case \cite{vo}.

%We saw that, regardless of the specific form of regularity conditions, regularity implies (\ref{Psi01-OZAMO}). Thus, taking into account the symmetry (\ref{OZAMO-symmetry}), all of the Weyl scalars turn to zero except for $\Psi_{2}$. Substitution of the explicit asymptotes (\ref{Psi01-OZAMO}) into the invariants (\ref{Iaxi}-\ref{Daxi}) gives
%\begin{equation}
%I=3\psi_{2}^{2}+O(N^2),\quad J=-\psi_{2}^{3}+O(N^2)
%\end{equation}
%and thus 
%\begin{equation}
%\Delta,L=O(N^2),\quad K=O(N^3),
%\quad M=O(N^4).
%\end{equation}
% In the general case, when $\Psi_{2}$ is separated from zero in the horizon limit, $I$ and $J$ do not vanish and the on-horizon boosted Petrov type is D. If, however, $\Psi_{2}$ vanishes at the horizon, then all the Weyl scalars tend to zero and the boosted Petrov type is O. This is the same situation as in the static case \cite{vo}.

\paragraph{Regular Petrov type (falling observer's frame).}

As $\Phi_{1,2}\to 0$, two of the pnds are aligned with the generator ($l_+$%
), so the metric is always algebraically special on the horizon, at least of
type II. In terms of invariants, indeed, using the explicit expressions (\ref%
{Iaxi}-\ref{Daxi}), we get (the same as in the OO frame) 
\begin{equation}
I^{(f)}=3\Phi_{2}^{2}+O(N^2),\quad J^{(f)}=-\Phi_{2}^{3}+O(N^2),
\end{equation}
where the $(f)$ superscripts denote that those invariants are calculated in
the FO frame, and thus 
\begin{equation}
\Delta^{(f)}=O(N^2).
\end{equation}

As opposed to OO frame, though, now $\Phi_{3,4}\neq 0$, so from (\ref{Inv-K}-%
\ref{Inv-M}) and (\ref{Phi0}-\ref{Phi4}) 
\begin{align}
K^{(f)}&\approx 8E^{3}\psi_{1} \big[2\psi_{1}^{2}-3\psi_{2}\psi_{0}\big]; \\
L^{(f)}&\approx 4E^{2}\; (\psi_{0}\psi_{2}-\psi_{1}^{2}); \\
M^{(f)}&\approx 48E^{4} \big[2\psi_{1}^{2}-3\psi_{2}\psi_{0}\big] \big[%
2\psi_{1}^{2}-\psi_{2}\psi_{0}\big].
\end{align}

The metric in the FO frame is of type $D$ when $K^{(f)}=M^{(f)}=0$, and this
happens if and only if (the other possible variants are specific cases) 
\begin{equation}
2\psi_{1}^{2}-3\psi_{2}\psi_{0}=0.
\end{equation}
In terms of original Weyl scalars this condition can be rewritten as 
\begin{equation}
X\equiv 2\Psi_{1}^{2}-3\Psi_{0}\Psi_{2}=o(N^2).
\end{equation}
The metric is of type III if $\psi_{2}$ vanishes but $\psi_{1}$ does not. It
is of type N if both of them vanish. Finally, it is of type O if and only if
additionally $\psi_{0}$ tends to zero.

\paragraph{Interpretation}

\begin{figure}[ht]
\centering
\begin{tikzpicture} [auto]
\pgfsetlinewidth{0.2ex}
	\tikzstyle{condshift} = [rectangle, draw, text centered, 
text width=4em, minimum height=2em, node distance=7em, inner sep=0pt, yshift=-1.5em]
	\node [condition, text width=5em, minimum height=2.7em] (X) 
	{$\displaystyle\frac{X}{N^2}\to 0$};
	\node [condshift, right of=X] (2) {$\psi_{2}\to 0$};
	\node [condshift, right of=2] (1) {$\psi_{1}\to 0$};
	\node [condshift, right of=1] (0) {$\psi_{0}\to 0$};
	\path [line] (X) -- node [near start] {no} (2);
	\path [line] (2) -- node [near start] {yes} (1);
	\path [line] (1) -- node [near start] {yes} (0);
	\node [type, below of=X, node distance=9em] (D) {D};
	\node [type, right of=D, node distance=7em] (II) {II};
	\node [type, right of=II, node distance=7em] (III) {III};
	\node [type, right of=III, node distance=5em] (N) {N};
	\node [type, right of=N, node distance=4em] (O) {O};
	\path [line] (X) -- node [swap] {yes} (D);
	\path [line] (2) -- node [swap] {no} (II);
	\path [line] (1) -- node [swap] {no} (III);
	\path [line] (0) -- node [swap, near end] {no} (N);
	\path [line] (0) -- node [near end] {yes} (O);
	\node [left of=D, node distance=4em] (FO) {Regular:};
	\node [below of=FO, node distance=2em] (OO) {Boosted:};
	\node [right of=OO, node distance=7.5em] (OOD)
	{$\underbrace{\qquad\qquad\qquad\qquad\;}
		_{\displaystyle\text{D}}$};
	\node [right of=OOD, node distance=15em] (OOO)
	{$\underbrace{\qquad
	\qquad\qquad\qquad\qquad}
		_{\displaystyle\text{O}}$};
\end{tikzpicture}

\caption{Flow diagram for determining the on-horizon regular (FO frame) and
boosted (OO frame) Petrov types.}
\label{TIKZ-F}
\end{figure}
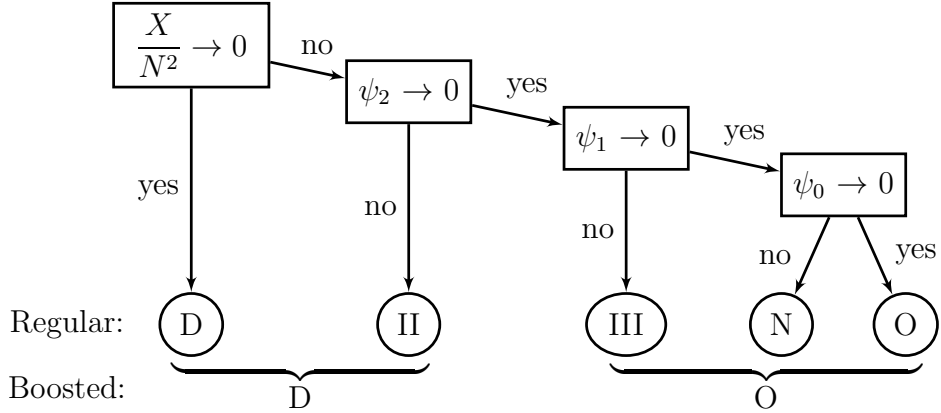

Summarizing all the variants considered so far, we can draw the flow diagram
shown on Fig. \ref{TIKZ-F}. The list of possible regular Petrov types given
the boosted Petrov type is D or O, given by the lowest two lines, reproduces
identically the table obtained for the generic static case in \cite{vo}.

We see that the result of calculating Petrov type can depend on the frame,
which can be expressed symbolically as 
\begin{equation}
\text{BPT}(r_{h})\neq \text{PT}(r_h),
\end{equation}
where PT stands for (regular) Petrov type, and BPT for the boosted Petrov
type, even though $\text{PT}(r)=\text{BPT}(r)$ hold for any $r>r_{h}$. In
other words, the operation of calculating Petrov type can be ``not
continuous'' on the horizon in the singular frame (if it is not continuous
in the regular frame either, then the horizon is an algebraically special
surface, -- see next section for more details).

The reason is of course the singular nature of the OO frame, that the pnds
in it are ``glued'' pairwise to the horizon's generator ($l_+$) and its time
reflection ($l_-$). This does not mean, however, that the structure itself
or its characteristic, $BPT$, is meaningless or incorrect. The result is the
consequence of symmetry and regularity. Formulation of the on-horizon
conditions of regularity in the OO frame contains important information
about space-time near the horizon (see eqs. (23)--(29) of \cite{v2} and eq.
(74) of \cite{tan}; also sec. IV of \cite{v2} and sec. VI of \cite{tan}, in
which the on-horizon structure of the Einstein tensor is analyzed). It is the OO frame, which is used often in physical applications. Say, the
results of calculations of the mean stress-energy tensor value of quantum
fields usually refer just to this frame, in particular the horizon value
(see. e.g. Sec. 11.3 of \cite{fn}).

Another point worth mentioning is the following. Regularity in the falling
frame allows one to restrict the on-horizon structure of the Einstein tensor
in the OO frame \cite{v2,tan}, but in the FO frame it remains essentially
arbitrary. In this regard, the degrees of freedom related to matter (encoded
in the Einstein tensor) behave differently to those of the free
gravitational field (contained in the Weyl tensor): the latter \emph{are}
constrained in \emph{both} frames. Even in the regular frame some of the Weyl scalars are shown to be zero, so that geometric structure of the Weyl tensor takes special form.

\subsection{On different singular boosts}

One may ask, whether it is possible to explain the correspondence between RPT and BPT on the horizon by the singular nature of the boost between the FO and OO frames only.  If this were true, our results would have nothing to do with the specific features of the horizon. However, more careful inspection shows that this is not so. Let us consider the procedure, which we will call the local singular boost here (the reason will become clear below), in more detail. Suppose there is some arbitrary point $P$, which may or may not belong to the horizon and in which spacetime is regular, i.e. Weyl scalar are finite. Let us make a boost at this point. In the appropriately chosen frame $\Phi_i$ transform into $\Psi_i$ according to (\ref{PhiS}). Now tend $x$ to infinity, making the boost singular. If we are to obtain a regular spacetime in the limit, we need $\Phi_{0,1}$ to be zero, which means that the boost is made along a multiple pnd. As $x\rightarrow \infty$, due to eq. (\ref{PhiS}) we have $\Psi_{3,4}\to 0$. Then in case $\Phi_2 \neq 0$ (i.e. Petrov type is II or D) the four roots in the new frame are $\lambda =0,0,\infty ,\infty $: the initial two pnds remain aligned, while the other two are merged together, and thus we obtain a spacetime of type D. If $\Phi_2 =0$, then all the components of the Weyl tensor vanish and we get a spacetime of type O. Thus the possible transformations of the PT are:
\begin{equation}
II,D\rightarrow D,
\quad III,N,O\rightarrow O. \label{tr}
\end{equation}
This corresponds to the 2nd and 3rd columns of Table \ref{Table} (see below). Therefore, it \textit{would seem} that the aforementioned lines of the Table are a trivial
consequence of the singular boost that has nothing to do with the horizons as such. However, this ``obvious'' conclusion is incorrect. 

The singular boost at the horizon, considered in this paper, is a quite different procedure both in physical and mathematical sense. Physically, it is realized naturally due to particle's motion near the horizon. Mathematically, it is decomposed into the following steps:
\begin{enumerate}
\item take a point $P$ parametrized by its coordinate distance to the horizon, or equivalently, the lapse function $N$;
\item make a boost with $x\sim 1/N$;
\item take the limit $N\to 0$, which implies $x\to \infty$.
\end{enumerate}
First, note that here we start from the frame that becomes singular in the considered limit, as opposed to the previous approach. This is a matter of convenience and problem statement. More importantly, this procedure actually takes into account not only the \emph{on-horizon values} of Weyl scalars, but their asymptotic behaviour in its vicinity, and depends on the latter crucially. 

In order to illustrate this, consider the following example. Let us start from the regular frame, in which Weyl scalars as functions of the lapse function $N$ are (the numbers are chosen to give simple algebra) 
\begin{equation}
\Phi_{0}=4N^2,\quad
\Phi_{1}=N,\quad
\Phi_{2}=1/6,\quad
\Phi_{3}=1,\quad
\Phi_{4}=1.  \label{ex1}
\end{equation}
As $\Phi_{0,1}$ at the horizon, where $N=0$, are zero, the Petrov type there is II. Now make a boost with $x=1/N$. According to eq. (\ref{PhiS}), in the new frame we obtain
\begin{equation}
\Psi _{0}=4,\quad
\Psi _{1}=1,\quad
\Psi _{2}=1/6,\quad
\Psi _{3}=N,\quad
\Psi _{4}=N^2.
\end{equation}
In the limit $N\to 0$ then the roots of (\ref{nullrot}) are $\lambda =1,3,\infty,\infty$, thus the type is II, not D! This is the consequence that $\Phi_{0,1}$ \emph{tend} to zero on the horizon, but are \emph{not identically} zero in the vicinity. 

With the same example, the \emph{local} singular boost with parameter $x\to\infty$ gives
\begin{align}
&\Phi _{0}=0,\quad \Phi _{1}=0,\quad \Phi _{2}=1/6,\quad\Phi _{3}=1,\quad  \Phi _{4}=1;\\
&\Psi _{0}=0,\quad \Psi _{1}=0,\quad \Psi _{2}=1/6,\quad\Psi _{3}=0,\quad  \Psi _{4}=0,
\end{align}
so in accordance with (\ref{tr}), we have type D. Thus the two procedures are clearly generally inequivalent. The asymptotic behavior of Weyl scalars does not matter to the local boost, but makes a big difference in our case.

The results look the same due to the the symmetry properties of the considered spacetime, which make $\Phi _{0,1}$ tend to zero fast enough (\ref{Phi0},\ref{Phi1}). Without them, the simple reasonings based on the properties of local singular boosts are insufficient.

It is of some interest therefore to consider and classify the possible relationships between RPT and BTP on the horizon when the symmetry requirements are
absent. Generalizing the above example (\ref{tr}), one can obtain the set of possible transformations from the regular to singular frame (or vice versa):
\begin{equation}
II,D\rightarrow II,D,\qquad II,N,O\rightarrow III,N,O. \label{Table-New}
\end{equation}
If the original type is I, the singular boost leads to diverging $\Psi_i$, so this case is absent. Besides regularity, the only effective link between the two frames turns out to be the  component $\Psi_2 =\Phi_2$, which either turns to zero or not in both frames simultaneously.

To summarize the contents of this section, there are two essentially different limiting procedures, which should not be mixed up. The local singular boost considered above is a purely mathematical operation in which one makes the boost ``manually'' in a given point. In contrast, in the problem we are dealing with in our paper, a different procedure is utilized. It is realized naturally in the physical setting: (almost \cite{ban}) any particle with finite energy that starts its motion in the outer region, reaches the horizon is such a way that the boost between its comoving frame and the static/orbital frame becomes singular on the horizon. The mathematical construction is also different and sensitive to the structure of spacetime in the vicinity, not only at the horizon itself. The results, rather surprisingly, if taking all of this into account, are the same only due to the assumed symmetry. Without it more possibilities exist.% and it is not clear whether there are some other considerations that could constrain them in the same way or not.

\section{Restrictions by the off-horizon Petrov types}
\label{sec:Off}
Suppose we have some exact solution, for which off-horizon, in its
arbitrarily small vicinity, the Petrov type is given. Is it possible then to
restrict the regular and boosted on-horizon Petrov types? Let us proceed to
answer this question starting from the more algebraically general
off-horizon metrics, and moving to the more special.

Off-horizon the classification is unambiguous, so it doesn't matter in what
frame the Petrov type is determined. In this section we will use the usual
OO frame, and take advantage of the simplified scheme for Petrov
classification given in section \ref{Petrov-axi-section}.

\begin{enumerate}
\item[I.] No simplifying conditions, so the result is generic: either the
boosted type is D and regular one is II or D, or the boosted type is O while
regular one is III, N or O.

\item[II.] Then $\pm \Psi_{1}=\Psi_{0}+3\Psi_{2}$, and the regularity
conditions (\ref{Psi01-OZAMO}) imply that on-horizon 
\begin{equation}
\Psi_{2}=\tfrac{1}{3}(\pm\Psi_{1}-\Psi_{0}) =O(N^{2})+O(N)=O(N)\to 0.
\end{equation}
Thus the boosted Petrov type can only be O. In the FO frame then the Petrov
type is III, N or O, depending on whether $\psi_{0}$ and $\psi_{1}$ tend to
zero or not (see Fig.~\ref{TIKZ-F}).

\item[D.] In this case $X\equiv 2\Psi_{1}^{2}-3\Psi_{0}\Psi_{2}=-\Psi_{0}^{2}
$. Then the regularity conditions (\ref{Psi01-OZAMO}) imply 
\begin{equation}
X=-\Psi_{0}^{2}=O(N^4)=o(N^2),
\end{equation}
so the Petrov type on-horizon is D or more special (as expected). If $%
\psi_{2}$ remains non-zero at the horizon, both the regular and boosted
types are $D$. Otherwise, if $\psi_{2}\to0$, then using the same regularity
condition we see that $\psi_{1}\to 0$, so the boosted Petrov type is O, and
the regular one is either O, if $\psi_{0}\to0$, or N otherwise (no type III).

\item[N.] In this case $\Psi_{0}=\Psi_{2}=\pm \Psi_{1}\neq 0$, so on-horizon 
\begin{equation}
\Psi_{1,2}=O(N^2)\quad\Rightarrow\quad \psi_{1,2}\to 0.
\end{equation}
The boosted Petrov type is necessarily O; the regular one is either O, if $%
\psi_{0}\to0$, or N otherwise.

\item[O.] All $\Psi_{i}$ are zero, so $X=0$ and $\psi_{1,2,3}=0$
identically, thus on-horizon both Petrov types are O.
\end{enumerate}

The results of this section are collected in Table \ref{Table-Off-On}, which
generalizes the one from \cite{vo} to the rotating case and relates the
on-horizon Petrov types to the off-horizon classification. Which of the
algebraic types is realized in the FO frame (second column) can be
determined from the flow diagram shown on Fig. \ref{TIKZ-F}. As mentioned
above, if we exclude possible but special cases of the off-horizon Petrov
types, marked with the asterisk in the table, in the generic case only types
I, D and O remain.

\begin{table}[!ht]
\centering
\begin{tabular}{||c|c|c||}
\hline
Off-horizon & \multicolumn{2}{|c||}{On-horizon Petrov types} \\ \cline{2-3}
Petrov type & Regular & Boosted $\phantom{\Big|}$ \\ \hline\hline
O & O & \multirow{3}{*}{O} \\ \cline{1-2}
N$^*$ & \multirow{2}{*}{N, O} &  \\ \cline{1-1}
\multirow{2}{*}{D} &  &  \\ \cline{2-3}
& D & \multirow{2}{*}{D} \\ \cline{1-2}
\multirow{2}{*}{I} & II, D &  \\ \cline{2-3}
& \multirow{2}{*}{\;III, N, O\;} & \multirow{2}{*}{O} \\ \cline{1-1}
II$^{*}$ &  &  \\ \cline{1-3}
\end{tabular}%
\caption{Possible off-horizon and on-horizon Petrov types. There are no type
III axisymmetric stationary metrics; variants marked by asterisks should not
be realized for generic ``real'' black holes.}
\label{Table-Off-On}
\end{table}

We see that the Petrov type calculated off-horizon, the one on horizon and
the boosted type on-horizon---in this very order---either become more
algebraically special, with multiplicities of pnds increasing, or the
algebraic type and multiplicities are preserved (for types D and O only).

The table \ref{Table-Off-On} implies, in particular, that the Weyl tensor at
the horizon may vanish in the OO frame but at the same time remain non-zero
in the FO one. To understand this seemingly paradoxical situation, let us
consider a simple analogy from special relativity. Suppose we have some
two-dimensional vector $l^{\mu}=N(a,b)$ where $N\ll1$ is a small parameter.
Then, by choosing sufficiently small $N$, we can make both components
arbitrarily small, so formally $l^{\mu}\rightarrow 0$ in this limit.
However, one can choose another Lorentz frame with the relative Lorentz
factor $\gamma\sim N^{-1}$. In this new frame both components of our vector
in the limit $N\to 0$ have the order $O(1)$, and remain separated from zero.
In this sense, the difference between two Petrov types under discussion is
mainly due to the kinematic effect.

The on-horizon (regular) Petrov type is generally more special than
off-horizon just because the horizon is a submanifold of the whole
space-time. If it differs, this means that the horizon is an algebraically
special surface: measurement of Petrov type distinguishes it from the bulk
space-time, contrary to the accepted notion that locally a horizon is
undetectable.

Thus, excluding the trivial O type case and specific special (not
``generic'') solutions, only Petrov type D metrics preserve their algebraic
structure at the horizon in both regular and singular frames, while type I
metrics must on-horizon become type D or II.

\section{Principal null directions}

\label{sec:principal}

\subsection{Generic case}

In this section and below we consider possible combinations of the Petrov
types using notation "A-B-C", where A is the off-horizon type, B is the
regular on-horizon type (in the FO frame) and C is the boosted on-horizon
type (as calculated in the OO frame).

Let us consider the asymptotic behavior of principal null directions
themselves. As mentioned above, they are given explicitly by $l_{+}^{\prime}$
of (\ref{nullrot-vec}), with $\lambda$ being the roots of quartic equation (%
\ref{nullrot}). In the OO frame, both on-horizon and not, due to symmetry
and regularity conditions (\ref{psi}) it is transformed to 
\begin{equation}  \label{OO-pndeq}
\psi_{0}N^{2}\lambda^{4}-4\psi_{1}N \lambda^{3} +6\psi_{2}\lambda^{2}
-4\psi_{1}N \lambda+\psi_{0}N^{2}=0,
\end{equation}
and in the FO frame due to (\ref{PhiS}) it takes the form 
\begin{equation}  \label{FO-pndeq}
x^{+2}N^2 \psi_{0}\lambda^{4} -x^{+1}N4\psi_{1}\lambda^{3}
+6\psi_{2}\lambda^{2} -4x^{-1}N\psi_{1}\lambda +x^{-2}N^{2}\psi_{0}=0.
\end{equation}
We will denote the roots of Eq. (\ref{OO-pndeq}) by $\lambda_{i}$ and the
roots of Eq. (\ref{FO-pndeq}) by $\lambda_{i}^{(f)}$. Note, that if $%
\lambda_{i}$ is a root of Eq. (\ref{OO-pndeq}), then 
\begin{equation}  \label{lambdas}
\lambda^{(f)}_{i}=\frac{\lambda_{i}}{x} =\frac{\lambda_{i}}{\gamma (1+v)} =%
\frac{N\;\lambda_{i}}{E(1+v)}
\end{equation}
is the root of Eq. (\ref{FO-pndeq}), and vice versa. Also if $\lambda_{i}$
is a root of Eq. (\ref{OO-pndeq}), then $1/\lambda_{i}$ is also its root.
Thus we only need to solve one of the two equations, and all the structure
in both frames is determined once we know two of the four roots of Eq. (\ref%
{OO-pndeq}), which are not inverse of each other.

Explicitly the pnds in each frame are given by Eq. (\ref{nullrot-vec}), with 
$\lambda=\lambda_{1,2,3,4}$. For small $\lambda_{i}\to 0$ we immediately
obtain a series by $\lambda$; for large $\lambda_{j}\to \infty$ we divide
the whole expression by the large factor $|\lambda_{j}|^{2}$ to obtain the
series by $\lambda_{j}^{-1}$: 
\begin{align}
\lambda_{i}\to0&\quad\Rightarrow\quad k_{i}=l_{+}+2\mathop{\rm{Re}} \big(%
m_{-}\lambda_{i}\big) +O(|\lambda_{i}|^2);  \label{pnd-small} \\
\lambda_{j}\to\infty&\quad\Rightarrow\quad k_{j}=l_{-}+2\mathop{\rm{Re}} %
\big(m_{+}\lambda_{j}^{-1}\big) +O(|\lambda_j|^{-2}).  \label{pnd-large}
\end{align}
Thus the asymptotics of $\lambda_{i}$ give us directly the "rate of
alignment" of the pnds along $l_{\pm}$.

It is important to remember that on-horizon $l_+$ is aligned with the
horizon's generator. Then Eqs. (\ref{pnd-small}) and (\ref{lambdas}) imply
that if on-horizon one of the pnds is aligned with $l_+$, so that $%
\lambda_{i}$ is small, then $\lambda^{(f)}_{i}$ is small too (even smaller)
and in the FO frame this pnd is aligned with the generator as well (but not
vice versa).

Let us assume at first, for simplicity, that all the functions $\psi_{0,1,2}$
are separated from zero in the vicinity of the horizon. Then in the horizon
limit $N\to 0$ the structure of Eq. (\ref{OO-pndeq}) implies that there are
two small roots $\lambda_{1,2}\sim N$ and two large roots $\lambda_{3,4}\sim
1/N$. Exact expressions can be given, but in order to see the asymptotic
structure it is more convenient to use the small parameter $N$ explicitly.
Expanding the exact solutions in powers of $N$, or alternatively, searching
for the roots in the form of series $\lambda=\Lambda_{-1}N^{-1}+\Lambda_{1}N
+\ldots$, and taking into account that in the horizon limit 
\begin{equation}
\frac{1}{x}=\frac{N}{2E}\big(1+O(N^2)\big),
\end{equation}
we obtain 
\begin{align}
&\lambda_{1,2}=\frac{y_\pm}{6\psi_2}\;N+O(N^3), & & \lambda_{3,4}=\frac{y_\pm%
}{\psi_0}\;\frac{1}{N}+O(N); \\
&\lambda_{1,2}^{(f)} =\frac{y_\pm}{6\psi_2}\;\frac{N^2}{2E}+O(N^4), & &
\lambda_{3,4}^{(f)} =\frac{y_\pm}{\psi_0}\;\frac{1}{2E}+O(N^2),
\end{align}
where 
\begin{equation}  \label{ypm}
y_{\pm}=2\psi_{1} \pm \sqrt{4\psi_{1}^{2}-6\psi_{0}\psi_{2}} =(2\Psi_{1}\pm%
\sqrt{2X})/N.
\end{equation}

Small $\lambda_{i}$ correspond to pnds which are aligned with $l_+$ (and
thus with the generator as well); large $\lambda_{i}$ correspond to those
aligned with $l_-$. Their explicit forms are obtained using (\ref{pnd-small}%
) and (\ref{pnd-large}): in the OO frame all four are aligned with $l_{\pm}$ 
\begin{align}
k_{1,2}&=l_{+} +\tfrac{1}{3}\mathop{\rm{Re}}\Big( \frac{y_\pm}{\psi_2} m_{-}%
\Big)\cdot N +O(N^2); \\
k_{3,4}&=l_{-} +2\mathop{\rm{Re}}\Big( \frac{\psi_0}{y_{\pm}}m_{+}\Big)\cdot
N +O(N^2),
\end{align}
and in the FO frame only two become aligned with $l_+$: 
\begin{align}
k^{(f)}_{1,2}&=l_{+} +\mathop{\rm{Re}}\Big( \frac{y_\pm}{\psi_2} m_{-}\Big)%
\cdot \frac{N^2}{6E} +O(N^4).
\end{align}

This picture works only if $\psi_{0}$, $\psi_{2}$ and $y_{\pm}$ are not zero
and do not tend to zero at the horizon, otherwise the derivation may be
invalid, as different orders of magnitude by $N$ become mixed up in the
process. If some of them are zero off-horizon, then off-horizon the
algebraic type may become more special; if they vanish on-horizon only, then
on-horizon it can be more special. None of this is an obstacle only for most
general metric, of type "I-II-D" in the notation introduced earlier. This
conforms with the conclusions of the previous section. All the more
algebraically special cases have to be considered separately, case by case.

\subsection{On-horizon boosted type D}

On-horizon in the OO frame the Petrov type is D unless $\psi_2$ is zero or
tends to zero. So here we consider what happens if $\psi_0$ or $y_{\pm}$ are
zero or tend to zero, while $\psi_2$ is assumed to be separated from zero.
Note, that off-horizon the Petrov type can be only I or D. If there are no
pairs of $\lambda_{i}$ with the same asymptotics, then it is automatically
type I.

\begin{enumerate}
\item Let $\psi_{0}=0$. Then Eq. (\ref{OO-pndeq}) is reduced to 
\begin{equation}
\lambda\big[2\psi_{1}N\lambda^{2}-3\psi_{2}\lambda +2\psi_{1}N\big]=0.
\end{equation}
and two of the roots in both frames are 
\begin{equation}
\lambda_{1,4}=0,\infty;\qquad \lambda_{1,4}^{(f)}=0,\infty,
\end{equation}
so one pnd is always aligned with $l_+$ and one is aligned with $l_-$.

\begin{enumerate}
\item If $\psi_{1}$ is separated from zero, the remaining two roots are 
\begin{align}
&\lambda_{2}=\frac{2\psi_1}{3\psi_2} N +O(N^3), & & \lambda_{3}=\frac{3\psi_2%
}{2 \psi_1} \frac{1}{N}+O(N) ,  \label{lambdaI-II-Da} \\
&\lambda_{2}^{(f)}=\frac{\psi_1}{3\psi_2} \frac{N^2}{2E}+O(N^4), & &
\lambda_{3}^{(f)}=\frac{3\psi_2}{2 \psi_1} \frac{1}{2E}+O(N^2).
\label{lambdaI-II-Db}
\end{align}

The pnds that become aligned with $l_{\pm}$ are 
\begin{align}
k_{2,3}&=l_{\pm} +\tfrac{2}{3}\mathop{\rm{Re}} \Big(\frac{\psi_{1}}{\psi_{2}}%
m_{\mp}\Big) \cdot N+O(N^2), \\
k_{2}^{(f)}&=l_{+} +\tfrac{2}{3}\mathop{\rm{Re}} \Big(\frac{\psi_{1}}{\psi_2}%
m_{-}\Big) \cdot\frac{N^{2}}{2E}+O(N^4).
\end{align}
The metric is of type "I-II-D".

\item If $\psi_1 \to 0$, instead of (\ref{lambdaI-II-Da}) and (\ref%
{lambdaI-II-Db}) we obtain 
\begin{align}
&\lambda_{2}=o(N),\quad\lambda_{3}=\frac{1}{o(N)}; \\
&\lambda_{2}^{(f)}=o(N^2), \quad\lambda_{3}^{(f)}=\frac{1}{o(1)},
\end{align}
so even though the asymptotics change, the algebraic structure is the same.

\item If $\psi_{1}=0$, then the roots are $\lambda_{1,2,3,4}=0,0,\infty,%
\infty$ and the type is "D-D-D".
\end{enumerate}

\item Let $\psi_{0}\to0$.

\begin{enumerate}
\item If $\psi_{1}$ is separated from zero, then $y_{+}\approx 4\psi_{1}$
and $y_{-}=o(1)$, so repeating the general derivation we get 
\begin{align}
&\lambda_{1}\sim N,\quad \lambda_{2}=o(N),\quad \lambda_{3}\sim \frac{1}{N}%
,\quad \lambda_{4}=\frac{1}{o(N)}; \\
&\lambda_{1}^{(f)}\sim N^2,\quad \lambda_{2}^{(f)}=o(N^2),\quad
\lambda_{3}^{(f)}\sim 1,\quad \lambda_{4}^{(f)}=\frac{1}{o(1)}.
\end{align}
Limits are the same and the type is "I-II-D", even though the asymptotic
structure may differ.

\item If $\psi_{1}=0$, then equation is bi-quadratic, $y_{\pm}\to 0$, and 
\begin{equation}
\lambda_{1}=-\lambda_{2}=o(N),\quad \lambda_{3}=-\lambda_{4}=\frac{1}{o(N)},
\end{equation}
therefore in both frames two pnds become aligned with $l_+$ and two with $l_-
$; the type is "I-D-D".

\item If $\psi_{1}\to 0$, the on-horizon types are the same as in the
previous case, so the type is again "I-D-D".
\end{enumerate}

\item $X=0$ but $\psi_{0}$ is separated from zero. Then the general
derivation holds with $y_{+}=y_{-}=2\psi_{1}\neq 0$, so two pairs of pnds
on-horizon are aligned. Off-horizon, however, checking the conditions $X=0$
and $\Psi_{0,2}\neq 0$ with Fig.~\ref{Petrov-Axi} shows that type D is
excluded. So the type is "I-D-D".

\item $X\to 0$ with $\psi_{0}$ separated from zero: on-horizon the structure
is the same, but off-horizon it is not restricted, thus the type can be
"I-D-D" or "D-D-D".
\end{enumerate}

\subsection{On-horizon boosted type O}

\label{Sec:On-horizon_type_O} As shown in Sec.~\ref{sec:regularity}, when $%
\psi_{2}$ is zero or tends to zero at the horizon, the on-horizon boosted
Petrov type is O, while the regular one can be O, N or III. Let us first
assume that $\Psi_{2}=0$. Then the equation (\ref{OO-pndeq}) is reduced to 
\begin{equation}
\xi\lambda^{4}-4\lambda^{3}-4\lambda +\xi =0,
\end{equation}
where 
\begin{equation}
\xi=\frac{\psi_0}{\psi_1}N,
\end{equation}
and the structure of its solutions depend only on asymptotic behavior of $\xi
$.

\begin{enumerate}
\item $\xi=0$. The exact solution is 
\begin{equation}
\lambda_{1,2,3,4}=0,\pm i,\infty, \quad\text{and}\quad
\lambda_{1,2,3,4}^{(f)}=0,O(N),O(N),\infty,
\end{equation}
so the type is "I-III-O".

\item $\xi\to 0$. Then the roots are 
\begin{equation}
\lambda_{1,2,3,4}=\frac{\xi}{4}+O(\xi^3), \pm i+O(\xi), \frac{4}{\xi}%
+O(\xi), \quad\text{and}\quad \lambda_{1,2,3,4}^{(f)} \sim \xi N, \pm i N, 
\frac{N}{\xi}.
\end{equation}

\begin{enumerate}
\item If $N/\xi=\psi_{1}/\psi_0$ is separated from zero (i.e. $%
\psi_{0}/\psi_{1}=O(1)$), then one of the pnds in the FO frame is still not
aligned with the generator and the type is again "I-III-O";

\item otherwise (i.e if $\psi_{1}/\psi_{0}\to 0$) all four $\lambda_{i}^{(f)}
$ tend to zero and the type is "I-N-O".
\end{enumerate}

\item $\xi$ is separated from zero and finite: there are four bounded
non-zero roots $\lambda_{i}$ and all $\lambda_{i}^{(f)}\sim N$, the type is
"I-N-O"; however, under some specific conditions, discussed in section \ref%
{Sec:IIN}, the off-horizon Petrov type can be II, thus "II-N-O".

\item $\xi\to \infty$: the roots are $\lambda_{1,2,3,4}=(-1)^{1/4}+O(%
\xi^{-1})$ and $\lambda_{i}^{(f)}\sim N$, same as above;

\item $\xi=\infty$: $\lambda_{1,2,3,4}=(-1)^{-1/4}$ and $\lambda_{i}^{(f)}%
\sim N$, same as above.
\end{enumerate}

%So, there are essentially only two distinct variants, which correspond to types "I-III-O" and "I-N-O". Recall however, that if all $\psi_{i}$ are zero, the type is "O-O-O". 

If $\psi_{2}\to 0$, we have two parameters, $N^{2}\psi_{0}/\psi_{1}$ and $%
N\psi_{1}/\psi_{0}$, each of which can be small, large or bounded, zero or
not. The full picture will also depend on whether $\lambda_i$ is $o(N^{-1})$
or not, which corresponds to $\tilde{\lambda}^{(f)}_{i}$ tending to zero or
not. There is little merit in sorting out all possible cases: each is easy
to analyze on its own. Here we will only give a couple more examples of
asymptotic behavior of $\psi_{0,1,2}$, illustrating the combinations of
off-horizon and on-horizon types not yet covered.

If $\psi_{0,1,2}\to 0$ but are separated from zero and $\Delta\neq 0$, then
the type is "I-O-O".

Let 
\begin{equation}
2\Psi_{1}^{2}=3\Psi_{0}\Psi_{2}-\Psi_{0}^{2},\quad \Psi_{0}\neq \Psi_{2},
\label{DXO}
\end{equation}
so that the off-horizon Petrov type is D. Then

\begin{enumerate}
\item if $\psi_{0,1}=0$, while $\psi_{2}\to0$ but is not identically zero,
then $\lambda_{1,2,3,4}=0,0,\infty,\infty$, two double pnds are aligned with 
$l_{\pm}$ and the type is "D-O-O". More generally, if $\psi_{0,1,2}\to 0$,
the algebraic type will be the same;

\item if $\psi_{2}\sim N^2$, $\psi_{1}\sim N$ but $\psi_{0}$ is separated
from zero, then all four roots $\lambda_{1,2,3,4}$ are finite, while $%
\lambda_{1,2,3,4}^{(f)}\sim N\to 0$ and the type is "D-N-O".
\end{enumerate}

Finally, let us consider the case of spacetimes with double circular pnds,
discussed in Sec.~\ref{Sec:IIN} . From the linear relation (\ref{PsiLinear})
and regularity conditions (\ref{Psi01-OZAMO}), (\ref{Psi34-FZAMO}) we see,
that for type D $\psi_{1}=0$ and $\psi_{2}=O(N^2)$ but is not zero; if $%
\psi_{0}\to 0$ then all $\Phi_{i}$ in (\ref{Phi0}-\ref{Phi4}) vanish and the
type is "D-O-O", otherwise $\lambda_{1,2,3,4}^{(f)}\sim N\to 0$ and the type
is "D-N-O". Likewise for type N: $\psi_{2}=O(N^2)$ and $\psi_{1}=O(N)$, so
the type is "N-O-O" if $\psi_{0}\to 0$ and "N-N-O" otherwise. 
%Both situations were not considered in previous sections.

In the same way for type II $\psi_{2}=O(N)$; if $\psi_{0,1}\to 0$ then the
type is "II-O-O". Excluding this case, there are two variants: "II-N-O" and
"II-III-O": three of the roots $\lambda_{i}^{(f)}$ are always small, so
three pnds are aligned with the generator. The remaining one is not aligned,
and the type is "II-III-O", when there exists one root $\lambda^{(f)}_{4}$
separated from zero, or equivalently, when there exists one root 
\begin{equation*}
\lambda_{3}=\lambda_{4}^{-1}  =\frac{N}{\lambda_{4}^{(f)}}=O(N)
\end{equation*}
of Eq. (\ref{roots34}). The last condition is equivalent to $%
\psi_{0}/\psi_{1}=O(1)$. The particular case $\psi_{0}\sim \psi_{1}$
corresponds to $\lambda_{4}^{(f)}$ being finite, otherwise $k_4^{(f)}$ is
aligned with $l_-$. 
%Only one sub-variant ($\psi_{2}=0$ and $\psi_{1}=O(N)\to 0$) was considered in section \ref{Sec:On-horizon_type_O}.

\subsection{Intermediate results}

By construction, in the horizon limit $l_+$ is aligned with the horizon
generator. The main results of this section can be summarized in the
following way. There are two main variants.

In the "generic" case Weyl tensor does not vanish in the horizon limit in
the OO frame and the pnds on-horizon are well-defined in both frames. Then
in the OO frame two pnds are aligned with the generator and two with $l_-$,
thus the boosted type is D. When we pass to the FO frame, the two pnds
aligned with the generator remain aligned with it, while one or both of
those previously aligned with $l_{-}$ can detach, but they do not align with
the generator, so the regular type remains D or becomes II. The off-horizon
type can be either I or D.

In the case the Weyl tensor vanishes on-horizon in the OO frame, the pnds in
the horizon limit remain well-defined (unless off-horizon type is trivial O)
and aligned in this limit with the generator and $l_-$; the Weyl tensor does
not have to vanish in general in the FO frame. When we pass to the FO frame,
two of the pnds aligned with the generator remain aligned (this is always
true), and one or both of the other two become aligned with it as well. The
off-horizon type can be either a) I or D, or b) II, D and N. Variant b)
corresponds to a special class of metrics, each possessing a double
(quadruple for type N) pnd associated with $\lambda=\pm 1$ and with a null
curve $r=const$ and $z=const$, discussed in Sec.~\ref{Sec:IIN}.

\section{Example: exotic regular ultra-extremal metrics}

\label{sec:example} As shown in \cite{tan}, there are two types of $p = 3$
ultraextremal regular metrics which do not obey the generic conditions of
"rigidity" (\ref{GenericRegCond}) and are called there "exotic".

The first exotic metric is "strange", with metric functions 
\begin{align}
N^{2}(r,z)&=\underline{\kappa_{3}}r^{3} +\underline{\kappa_{4}}r^{4}
+\kappa_{5}(z)r^{5}+\ldots;  \label{strange-b} \\
\omega(r,z)&=\underline{\omega_{H}} +\underline{\omega_{2}}r^{2}
+\omega_{3}(z)r^{3}+\ldots; \\
g_{\phi\phi}(r,z)&=\underline{g_{\phi H}} +g_{\phi 1}(z)r+\ldots; \\
g_{zz}(r,z)& =C_{0}\cdot(\kappa_{5}^{\prime})^{2} +g_{z1}(z)r+\ldots.
\label{strange-e}
\end{align}
The underlined quantities are constants here and below, $C_{0}=const$, while 
$\omega_{1}$ is absent. Note that for a generic regular metric with triple
horizon one would have $N^2 \sim r^3$ and thus $\kappa_{5}=const$.

The second exotic metric is even more strange, and thus "weird": 
\begin{align}
N^{2}(r,z)&=\underline{\kappa_{3}}r^{3} +\kappa_{4}(z)r^{4}
+\kappa_{5}(z)r^{5}+\ldots;  \label{weird-b} \\
\omega(r,z)&=\underline{\omega_{H}} +\omega_{3}(z)r^{3}+\omega_{4}(z)r^{4}+%
\ldots; \\
g_{\phi\phi}(r,z)&=\underline{g_{\phi H}} +\underline{g_{\phi 1}}r+g_{\phi
2}(z)r^{2}+\ldots; \\
g_{zz}(r,z)&=C_{1}\cdot(\kappa_{4}^{\prime})^{2} \Big[1+\Big( 2\frac{%
\kappa_{5}^{\prime}}{\kappa_{4}^{\prime}} -\frac{\kappa_4}{\kappa_3}+C_{2}%
\Big)r\Big] +g_{z 2}(z)r^{2}+\ldots,  \label{weird-e}
\end{align}
where $C_{1,2}$ are constants. Note the absence of both $\omega_{1}$ and $%
\omega_{2}$. In both cases the expansion of $A(r)$ is generic: 
\begin{equation}  \label{ultra-A}
A(r)=\underline{\alpha_{3}}r^{3} +\underline{\alpha_{4}}r^{4}+\ldots.
\end{equation}

In this section, we analyze the structure and Petrov type of these two
metrics, and show that they not only look strange, but are algebraically
special at the horizon. We start from curvature and Ricci tensors, look at
scalar invariants -- Ricci $R$ and Kretchmann $Kr$ scalars, and 
\begin{equation}  \label{R2}
R_{2}=R_{\mu\nu}R^{\mu\nu}-\tfrac{1}{4}R^{2}
\end{equation}
(this is the traceless part of Ricci tensor squared), -- check out next
order differential invariants, finally calculate the Weyl invariants and
determine both the regular and boosted Petrov types. As the results for the
two are very similar, first we investigate the strange metric, and then
state what is different for the weird one.

\subsection{Riemann tensor, contractions and invariants}

First we consider the OO frame. Curvature tensor components are generally
separated from zero, but the Ricci tensor turns to zero on the horizon: 
\begin{align}
R_{(0)(0)}\approx -R_{22}&\sim r; \\
R_{(1)(1)}\approx R_{33}&\sim r; \\
R_{(0)(1)}\sim R_{23}&\sim r^{3/2}; \\
R_{(0)(2)},R_{(0)(3)},R_{(1)(2)},R_{(1)(3)}&=0.
\end{align}
The last two lines are due to regularity and symmetry respectively, and they
hold for the generic metric also. The first two lines are true for exotic
metrics only.

The invariants are 
\begin{equation}  \label{strange-inv}
R\sim r;\quad R_{2}, Kr \sim r^{2}.
\end{equation}
Thus the space-time is Ricci flat in the horizon limit and all algebraic
invariants of the curvature tensor turn to zero on it. It is then worth
looking at differential invariants, for example 
\begin{equation}  \label{DiffInv}
R_{;\mu}R^{;\mu},\quad R_{\mu\nu;\lambda}R^{\mu\nu;\lambda},\quad
R_{\mu\nu\rho\sigma;\lambda} R^{\mu\nu\rho\sigma;\lambda}.
\end{equation}
All of these three invariants also turn out to vanish on the horizon. In the
FO frame the curvature and Ricci tensor's components do not vanish in the
horizon limit, but all the invariants obviously do. This is the consequence
of the fact that, as discussed in \cite{v}, when invariants of zeroth order
all vanish at some point, it is nontrivial to find non-vanishing
higher-order invariants.

\subsection{Weyl scalars and Petrov types}

The Weyl scalars are 
\begin{align}
\Psi_{0,4}&\sim r^{3}, \\
\Psi_{1,3}&\sim r^{3/2}; \\
\Psi_{2}&\sim r.
\end{align}
The first two lines are necessary due to regularity. The difference from the
generic case is that $\Psi_{2}\to 0$, however $\psi_{1}$ and $\psi_{0}$
still do not tend to zero: 
\begin{equation*}
\psi_{0}\nrightarrow 0,\quad \psi_{1}\nrightarrow 0,\quad
\psi_{2}\rightarrow 0.
\end{equation*}
Thus the on-horizon boosted Petrov type is O, and the regular one is III.
Off-horizon it can be then of either type II or I. Of course, the algebraic
type can be more special if additional constraints on the metric are
satisfied.

As shown in \cite{v}, the simplest invariants that can be non-zero for type
III or N metrics with twist or expansion are 
\begin{align}
I_{1}&=C^{\alpha\beta\gamma\delta;\epsilon} C_{\alpha\mu\gamma\nu;\epsilon}
C^{\lambda\mu\rho\nu;\sigma} C_{\lambda\beta\rho\delta;\sigma} =\big(48\rho 
\bar{\rho} \Psi_3 \bar{\Psi}_3\big)^2;  \label{DiffInv1} \\
I_{2}&=C^{\alpha\beta\gamma\delta;\epsilon\phi}
C_{\alpha\mu\gamma\nu;\epsilon\phi} C^{\lambda\mu\rho\nu;\sigma\tau}
C_{\lambda\beta\rho\delta;\sigma\tau} =\big(48\rho^2 \bar{\rho}^2 \Psi_4 
\bar{\Psi}_4\big)^2,  \label{DiffInv2}
\end{align}
where $\rho=-(l_{+})_{\mu;\nu}m_{+}^{\mu}m_{-}^{\nu}$ is one of the spin
coefficients. However, in our case $\Psi_{3,4}\to 0$ and it can be checked
that $\rho\sim r$ in the OO frame, so both of these vanish on the horizon
too.

\subsection{Intrinsic geometry and topology of the horizon}

The on-horizon 2-dimensional line element 
\begin{equation}
dl_{2}^{2}=g_{\phi\phi} d\phi^{2}+g_{zH}(z)dz^{2} =d\tilde{\phi}^{2}+d\tilde{%
z}^{2},
\end{equation}
can be reduced to Euclidean form by simple coordinate transformation to $%
\tilde{\phi}(\phi)$ and $\tilde{z}(z)$, as $g_{\phi\phi}$ is a constant and $%
g_{zH}$ depends only on $z$. Thus the metric is flat, and its topology
variants are limited to Euclidean plane, cylinder or torus (see \cite%
{Katanaev}). Solutions with such horizons are known \cite{top1,top2,top3}.

\subsection{The weird metric}

All curvature tensor components tend to zero, with the component that does
so the slowest being 
\begin{equation}
R_{(0)(2)(0)(2)}=3\alpha_{3}r+O(r^2),
\end{equation}
where $\alpha_{3}$ is the leading coefficient in the expansion of $A(r)$ (%
\ref{ultra-A}), and all the other components are $O(r^2)$, so the metric is
asymptotically flat on the horizon. Thus all the invariants turn to zero,
with asymptotes 
\begin{align}
R&\approx -6\alpha_{3}r; \\
R_{2}\approx Kr &\approx 36\alpha_{3}^{2}r^{2}.
\end{align}
The higher-order invariants (\ref{DiffInv}), (\ref{DiffInv1}) and (\ref%
{DiffInv2}) also tend to zero.

All the Weyl scalars have the same asymptotes as in the previous case, with 
\begin{equation*}
\Psi_{2}=-\frac{\alpha_3}{2}r +O(r^2)
\end{equation*}
so on-horizon the boosted Petrov type is O andthe regular one is III; off
horizon it is of type I or II.

The intrinsic geometry is flat, so the topological variants are again
limited to Euclidean plane, cylinder and torus.

\section{Discussion and conclusion}

\label{sec:discussion}

We have considered here the frames of two classes of observers: the usual zero angular momentum observers on circular orbits, or just orbital observers (OO), and the falling observers (FO), that cross the horizon with finite proper acceleration. The true Petrov type (RPT) of the metric at the horizon is determined in the FO frame, which is regular. The Petrov type calculated formally in the OO frame, which is related to the FO frame by a singular Lorentz boost, with $\gamma \sim 1/N$, can be different and more special. We call it the boosted Petrov type (BPT) on the horizon. The main results on the correspondence between possible values of RPT and BPT on horizon and the Petrov type off-horizon are given in table \ref{Table} (it is essentially table \ref{Table-Off-On} read from right to left).

\begin{table}[!ht]
\centering
\begin{tabular}{||c|c|l||}
\hline\hline
\multicolumn{2}{||c|}{On-horizon Petrov types (PT)\;} & Off-horizon \\ \cline{1-2}
Boosted PT & Regular PT & Petrov type$\phantom{\Big|}$ \\ \hline\hline
\multirow{2}{*}{D} & II & \; I \\ \cline{2-3}
& D & \; I, D \\ \hline
\multirow{3}{*}{O} & III & \; I, II$^*$ \\ \cline{2-3}
& N & \; I, II$^*$, D, N$^*$ \\ \cline{2-3}
& O & \; I, II$^*$, D, N$^*$, O \\ \hline\hline
\end{tabular}%
\caption{Possible combinations of on-horizon boosted PT (as calculated in the singular frame of the
observer on a circular orbit, regular PT (as calculated in the frame of one falling through
the horizon), and the algebraic type of space-time near the horizon. Asterisk marks off-horizon types not realized
in the ``generic'' case.}
\label{Table}
\end{table}

The constraints on on-horizon algebraic structure in both frames are due to the combination of i) existence of a horizon, ii) the demand that space-time is regular on it, and iii) the symmetries. For a falling observer the regularity conditions force two of the four principal null directions to be aligned with the generator of the horizon. Making a singular Lorentz boost to the frame of the orbital observer, we align the remaining two principal null directions along another direction, $l_-$. Thus the boosted Petrov type can be only D (or trivially O).

The horizon limit in the OO and FO frames bears very different physical
meaning. For the falling observer, this limiting procedure is realized by
the observer actually moving along its worldline and across the horizon, in
finite proper time. For OO this is the formal limit taken by changing the
observers of the given class up to the point when they become light-like, as
each of them orbits the horizon at constant $r$. Even though this limit is
well-defined, the resulting algebraic structure is not actually ``observed''
by any single observer.

We showed that the corresponding Lorentz boost that relates the two frames and becomes singular in the horizon limit is a different mathematical structure than the ``trivial'' singular boost defined at a point. The table of correspondence between PT and BPT, summarized in Table \ref{Table}, turns out to be the same only due to additional symmetry. If symmetry assumptions are relaxed, the list of correspondence (see Eq. \ref{Table-New}) looks different.

In the general case one can see that columns II and III of Table \ref{Table} differ. This means that the horizon turns out to be an algebraically special
surface, on which the Petrov type of the metric is different from the one in the surrounding space-time, which contradicts the belief that locally a
horizon is always undetectable. Space-times of Petrov type D, however, are an exception from this: as seen from the table, their algebraic type can be
preserved on the horizon.

We have also analyzed the process of principal null directions' alignment
with each other and with the horizon's generator, and found their asymptotes
in terms of series by the lapse function $N$ in all cases of interest. It
has also been shown, that all axisymmetric stationary space-times of
off-horizon types II and N, as well as a subclass of type D metrics, have
peculiar structure, possessing a double (quadruple for type N) pnd
associated with congruences of null curves $r=const$ and $z=const$.

Two reservations are in order. For extremal horizons there exist so-called
``critical'' observers, for which the proper time required to reach the
horizon is infinite. They are especially important for the so-called BSW
effect, which consists of indefinite growth of the energy in the centre of
mass frame of two particles at their collision near a horizon \cite{ban}.
The essential feature of critical observers is that their local velocity
measured by an OO does not tend to that of light, in contrast to all other
observers who reach the horizon \cite{k}. As the velocity is finite, there
is no singular Lorentz boost, and there is no disagreement between them and
orbital observers. Thus what really matters is whether an observer crosses
the horizon or not. There is also another exception when the observer does
cross the horizon but, nonetheless, no disagreement between the two kinds of
observers is expected. This happens if an observer passes through the
bifurcation point (that is relevant for the BSW effect inside black holes 
\cite{bifurk}). In this case, the local velocity of such a FO measured by the
analogue of the OO observer is also finite.

We have analyzed the two exotic regular ultra-extremal metrics found in \cite{tan} and showed that they are algebraically special in the horizon limit:
in the OO frame, one is Ricci flat and the other is flat, with algebraic and differential invariants vanishing in both cases. In the FO frame, however,
both are of Petrov type III on horizon, which is more general than their boosted Petrov type (calculated in the OO
frame). This conforms to the result for generic metrics, that BPT is (the same or) more algebraically special than PT.

The case of exotic metrics is interesting also in the following sense. The
existence of black hole horizon is usually thought of as a feature inherent
to strong gravitational field, absent in the weak field approximation.
However, those exotic regular metrics provide examples of horizons, on which
all zero order curvature invariants vanish. In this respect, it is worth
reminding that there exist metrics with nonzero Riemann tensor for which all
such invariants vanish everywhere \cite{v}. In this context, our results can
be viewed as a counterpart to this class of solutions with the reservation
that invariants now vanish not everywhere but on the horizon only.

Furthermore, the intrinsic geometry of the considered exotic metrics is
flat, which reduces the topological variants to plane, cylinder and torus.
Black hole solutions with such horizons are known \cite{top1,top2,top3}, but
whether the exotic metrics discussed correspond to any of them or to
acceleration horizons, remains an open question.

The local character of the effect implies that it is a feature of light-like
apparent horizons, and may still be present in the dynamical case when those
do not coincide with event horizons or the latter might not even exist. The
current study, however, relies heavily on the presence of symmetry. Careful
analysis of the dynamical situation is needed in order to say more on this
matter.

\begin{acknowledgements}
We thank Vojtech Pravda for the interest to this work and helpful discussion. The work of I.T. was supported in part by the Joint DFFD-RFBR Grant \#F40.2/040.
\end{acknowledgements}


\begin{thebibliography}{99}
\bibitem{p} Petrov, A. Z.: Classification of spaces defined by gravitational
fields. Uch. Zapiski Kazan Gos. Univ. 144, 55 (1954)

\bibitem{fn} Frolov V.P. and Novikov I. D.: Black Hole Physics: Basic
Concepts and New Developments. Kluwer Academic, Boston (1998)

\bibitem{m} Thorne, K.S., Price, R.H., Macdonald, D.A. (eds): Black Holes:
The Membrane Paradigm. Yale University Press, London (1986)

\bibitem{qbh} Lemos, J.P.S., Zaslavskii, O.B.: The angular momentum and mass
formulas for rotating stationary quasi-black holes. Phys. Rev. D 79, 044020
(2009); \href{http://arxiv.org/abs/0901.3860}{arXiv:0901.3860 [gr-qc]}

\bibitem{ban} Ba\~{n}ados, M., Silk, J., West, S. M.: Kerr black holes as
particle accelerators to arbitrarily high energy. Phys. Rev. Lett. 103,
111102 (2009)%
; \href{http://arxiv.org/abs/0909.0169}{arXiv:0909.0169 [hep-ph]}

\bibitem{k} Zaslavskii, O.B.: Acceleration of particles by black holes:
kinematic explanation. Phys. Rev. D 84, 024007 (2011)%
; \href{http://arxiv.org/abs/1104.4802}{arXiv:1104.4802 [gr-qc]}

\bibitem{bifurk} Zaslavskii, O.B.: Acceleration of particles near the inner
black hole horizon. Phys. Rev. D 85, 024029 (2012); 
 \href{http://arxiv.org/abs/1110.5838}{arXiv:1110.5838 [gr-qc]}; 
Zaslavskii, O.~B.: Horizon bifurcation surface as particle accelerator. Int.
Journ. Mod. Phys. D. 22, 1350044 (2013) 
; \href{http://arxiv.org/abs/1203.5291}{arXiv:1203.5291 [gr-qc]} 

\bibitem{v2} Medved~A.J.M., Martin~D. and Visser~M.: Dirty black holes:
Symmetries at stationary non-static horizons. Phys. Rev. D. 70, 024009
(2004) %; \href{http://arxiv.org/abs/gr-qc/0403026}{arXiv:gr-qc/0403026}

\bibitem{v3} Medved~A.J.M.: Symmetries at stationary Killing horizons. Gen.
Rel. Grav. 37, 1947-1956 (2005) 
; \href{http://arxiv.org/abs/gr-qc/0410103}{arXiv:gr-qc/0410103}

\bibitem{tan} Tanatarov I.V., Zaslavskii, O.B.: Dirty rotating black holes:
regularity conditions on stationary horizons. Phys. Rev. D 86, 044019 (2012) 
; \href{http://arxiv.org/abs/1206.2580}{arXiv:1206.2580 [gr-qc]}

\bibitem{conf} Kunduri H., Lucietti J., Reall H.: Near-horizon symmetries of
extremal BHs. Class. Quant. Grav. 24, 4169-4190 (2007)
 \href{http://arxiv.org/abs/0705.4214}{arXiv:0705.4214 [hep-th]};
  Guica M., Hartman T., Song W. and Strominger A.: The Kerr/CFT
Correspondence. Phys. Rev. D 80, 124008 (2009)
 \href{http://arxiv.org/abs/0809.4266}{arXiv:0809.4266 [hep-th]};
  Mei J.: Conformal symmetries of the Einstein-Hilbert action on horizons of
stationary and axisymmetric black holes. Class. Quant. Grav. 29, 095020
(2012) \href{http://arxiv.org/abs/1108.3841}{arXiv:1108.3841 [hep-th]};
 Ortin T., Shanbazi C.: A note on the hidden conformal structure of
non-extremal black holes. Phys. Lett. B 716, 231-235 (2012)
 \href{http://arxiv.org/abs/1204.5910}{arXiv:1204.5910 [hep-th]}.

\bibitem{carlip} Carlip S.: Effective Conformal Descriptions of Black Hole
Entropy: A Review. AIP Conf. Proc. 1483, 54-62 (2012);
 \href{http://arxiv.org/abs/1207.1488}{arXiv:1207.1488 [gr-qc]}
%Proceedings of the Sixth International School on Field Theory and Gravity, Petropolis, Brazil, 2012

\bibitem{d} Papadopoulos, D., Xanthopoulos, B.C.: Local black holes are type
D on the horizon. Nuovo Cimento B 83, 115 (1984)

\bibitem{vo} Pravda, V., Zaslavskii, O.B.: Curvature tensors on distorted
Killing horizons and their algebraic classification. Class. Quant. Grav. 22,
5053 (2005); \href{http://arxiv.org/abs/gr-qc/0510095}{arXiv:gr-qc/0510095}

\bibitem{boost} Aichelburg, P.C., Sexl, R.U.: On the gravitational field of
a massless particle. Gen. Relat. Grav. 2, 303 (1971)

\bibitem{exact} Stephani, H., Kramer, D., Maccallum, M., Hoenselaers, C.,
Herlt, E.: Exact solutions of Einstein's field equations. CUP, Cambridge
(2003)

\bibitem{72} Bardeen J., Press W.H., and Teukolsky S.A., Astrophys. J. 178, 347 (1972)

\bibitem{bsw-force} Tanatarov I.V, Zaslavskii O.B.: Ba\~{n}ados-Silk-West effect with nongeodesic particles: extremal horizons. \href{http://arxiv.org/abs/1307.0034}{arXiv:1307.0034 [gr-qc]}

\bibitem{Gr} Griffiths, J.B., Podolsky, J.: Exact space-times in Einstein's
General Relativity. CUP (2009)

\bibitem{TaylorWheeler} Taylor, E.F., Wheeler, J.A.: Exploring black holes:
Introduction to general relativity, F-17. AW Longman (2000)

\bibitem{el} Ellis, G.F.R., Schmidt, B.G.: Singular space-times. Gen.
Relativ. Gravit. 8, 915 (1977)

\bibitem{Ashtekar} Ashtekar, A., Fairhurst, S., Krishnan, B.: Isolated
horizons: hamiltonian evolution and the first law. Phys Rev D. 62, 104025
(2000); \href{http://arxiv.org/abs/gr-qc/0005083}{arXiv:gr-qc/0005083}

\bibitem{PPO2007} Pravda, V., Pravdova, A., Ortaggio, M.: Type D Einstein
spacetimes in higher dimensions. Class. Quant. Grav. 24, 4407 (2007); \href{http://arxiv.org/abs/0704.0435}{arXiv:0704.0435 [gr-qc]}

\bibitem{v} Bi\v{c}\'{a}k, J., Pravda, V.: Curvature invariants in type N
spacetimes. Class. Quantum Grav. 15, 1539 (1998)
 \href{http://arxiv.org/abs/gr-qc/9804005}{arXiv:gr-qc/9804005};
Pravda, V.: Curvature invariants in type-III spacetimes. Class. Quantum
Grav. 16, 3321 (1999) 
\href{http://arxiv.org/abs/gr-qc/9906121}{arXiv:gr-qc/9906121}

\bibitem{Katanaev} Katanaev, M.O., Kl\"{o}sch, T., Krummer, W.: Global
properties of warped solutions in General Relativity. Annals of Physics 276,
191 (1999); \href{http://arxiv.org/abs/gr-qc/9807079}{arXiv:gr-qc/9807079}

\bibitem{top1} Brill, D.R., Louko, J., Peld\'{a}n. P.: Thermodynamics of
(3+1)-dimensional black holes with toroidal or higher genus horizons. Phys.
Rev. D 56, 3600 (1997); \href{http://arxiv.org/abs/gr-qc/9705012}{arXiv:gr-qc/9705012}

\bibitem{top2} Vanzo, L.: Black holes with unusual topology. Phys. Rev. D
56, 6475 (1997); \href{http://arxiv.org/abs/gr-qc/9705004}{arXiv:gr-qc/9705004}

\bibitem{top3} Lemos, J.P.S.: Gravitational collapse to toroidal,
cylindrical and planar black holes. Phys. Rev. D 57, 4600 (1998); \href{http://arxiv.org/abs/gr-qc/9709013}{arXiv:gr-qc/9709013}
\end{thebibliography}
\end{document}